\documentclass[fleqn,usenatbib]{mnras}

\usepackage[T1]{fontenc}
\usepackage{ae,aecompl}

\usepackage{graphicx}	
\usepackage{amsmath}	
\usepackage{amssymb}	
\usepackage{multirow}

\newcommand{\msun}{{\rm M}_{\odot}}
\DeclareMathAlphabet{\mathpzc}{OT1}{pzc}{m}{it}

\usepackage{newtxtext,newtxmath}

\title[GOGREEN Transition Galaxies]{The GOGREEN survey: Transition Galaxies and The Evolution of Environmental Quenching}

\author[Karen McNab et al.]{
\newauthor
Karen McNab$^{1,2}$\thanks{E-mail: klmcnab@uwaterloo.ca},
Michael L. Balogh$^{1,2}$,
Remco F. J. van der Burg$^{3}$,
Anya Forestell$^{1,2}$,
\newauthor 
Kristi Webb$^{1,2}$, 
Benedetta Vulcani$^{4}$,
Gregory Rudnick$^{5}$,
Adam Muzzin$^{6}$,
M. C. Cooper$^{7}$,
\newauthor 
Sean McGee$^{8}$,
Andrea Biviano$^{9,10}$,
Pierluigi Cerulo$^{11}$,
Jeffrey C. C. Chan$^{12}$,
\newauthor
Gabriella De Lucia$^{13}$, 
Ricardo Demarco$^{14}$,
Alexis Finoguenov$^{15}$,
Ben Forrest$^{12}$,
\newauthor
Caelan Golledge$^{5}$,
Pascale Jablonka$^{16,17}$,
Chris Lidman$^{18,19}$,
Julie Nantais$^{20}$,
\newauthor
Lyndsay Old$^{21}$, 
Irene Pintos-Castro$^{22,23}$, 
Bianca Poggianti$^{4}$,
Andrew M. M. Reeves$^{1,2}$,
\newauthor
Gillian Wilson$^{12}$,
Howard K. C. Yee$^{23}$,
Dennis Zaritsky$^{24}$\\
\\
Author affiliations are listed at the end of the paper
}

\date{Accepted XXX. Received YYY; in original form ZZZ}

\pubyear{2021}

\begin{document}
\label{firstpage}
\pagerange{\pageref{firstpage}--\pageref{lastpage}}
\maketitle

\begin{abstract}
We measure the rate of environmentally-driven star formation quenching in galaxies at $z\sim 1$, using eleven massive ($M\approx 2\times10^{14}\,\mathrm{M}_\odot$) galaxy clusters spanning a redshift range $1.0<z<1.4$ from the GOGREEN sample.  
We identify three different types of transition galaxies:  "green valley" (GV) galaxies identified from their rest-frame $(NUV-V)$ and $(V-J)$ colours; "blue quiescent" (BQ) galaxies, found at the blue end of the quiescent sequence in $(U-V)$ and $(V-J)$ colour; and spectroscopic post-starburst (PSB) galaxies. 
We measure the abundance of these galaxies as a function of stellar mass and environment.  For high stellar mass galaxies ($\log{M/\mathrm{M}_\odot}>10.5$) we do not find any significant excess of transition galaxies in clusters, relative to a comparison field sample at the same redshift.  It is likely that such galaxies were quenched prior to their accretion in the cluster, in group, filament or protocluster environments.
For lower stellar mass galaxies ($9.5<\log{M/\mathrm{M}_\odot}<10.5$) there is a small but significant excess of transition galaxies in clusters, accounting for an additional $\sim 5-10$ per cent of the population compared with the field. 
We show that our data are  consistent with a scenario in which 20--30 per cent of low-mass, star-forming galaxies in clusters are environmentally quenched every Gyr, and that this rate slowly declines from $z=1$ to $z=0$.  While environmental quenching of these galaxies may include a long delay time during which star formation declines slowly, in most cases this must end with a rapid ($\tau<1$\,Gyr) decline in star formation rate.

\end{abstract}




\section {Introduction}
 
In a $\Lambda$CDM universe, structure evolution is driven by gravitational growth of dark matter haloes. Most of the elements necessary for galaxy formation under these circumstances have been understood for some time \citep{white1991galaxy}.  In particular, the growth of galaxies is dominated by star formation, mergers and feedback, which largely decouple them from the growth of dark matter structures \citep[e.g.][]{Behroozi13}. We can directly observe this galactic growth through the evolution of the stellar mass function (SMF), which shows that much of the stellar mass in the local universe was already assembled by $z\sim 2$  \citep[e.g.][]{2006A&A...459..745F,2007A&A...476..137A,2009ApJ...707.1595D,2012A&A...538A..33S,ultravista}.

The star-formation rate (SFR) of star-forming galaxies is nearly proportional to stellar mass, at all redshifts to at least $z=2.5$ \citep{salim2007uv,2007ApJ...660L..43N,2009ApJ...705L..67P,2012ApJ...754L..29W}, yet the SMF of these galaxies evolves only modestly with time \citep[e.g.][]{ultravista,2021MNRAS.503.4413M}.  In contrast, the abundance of non-star-forming (passive, or quiescent) galaxies builds up rapidly with time.  This indicates that there is a transition --- quenching --- of the galaxies from the blue cloud to the red sequence in colour--magnitude space as they stop forming stars \citep[e.g.][]{2004ApJ...608..752B,2007ApJ...654..858B,2007ApJ...665..265F}. The bimodality in the SFR and colour distributions of galaxies implies that this transition must be fairly rapid \citep[e.g.][]{wetzel2012galaxy}.
This transition rate is strongly dependent on stellar mass, and galaxies with high masses are more likely to be quenched \citep[e.g.][]{takamiya1995photoelectric,strateva2001color,ultravista}.  However, the
fraction of quiescent galaxies also depends on their environment \citep[e.g.][]{Lewis02,Gomez03,baldry2006galaxy} such that, at fixed stellar mass, galaxies in dense environments are more likely to be quiescent. 
This suggests that additional environmental quenching processes are contributing to the build up of the red sequence in clusters. 
Empirically, the additional environmental quenching at $z=0$ is separable from the stellar mass dependence \citep{baldry2006galaxy,Peng2010}, though this may not be true at $z\gtrsim 0.7$ \citep[e.g.][]{Balogh16,2017ApJ...847..134K}.

The intermediate region between the star-forming and quiescent phases of galaxy growth can provide constraints on the nature of galaxy transformations and the rate at which galaxies undergo them.
Such transition galaxies have often been used to trace the quenching rate over time and as a function of stellar mass \citep{tran2004field,pattarakijwanich2016evolution,wild2016evolution,wild2020formation,socolovsky2018enhancement}. 
The slow, undisturbed fading of star formation in galaxies may account for most of the quiescent population \citep[e.g.][]{2013ApJ...770...64G}, though there is also evidence that some galaxies undergo a more rapid transformation 
\citep[e.g.][]{belli2019mosfire}.   
The redshift evolution of these quenching pathways remains relatively unconstrained. This is in part due to differing definitions used to select transitioning galaxies, and to different assumptions about their parent populations
\citep[e.g.][]{belli2019mosfire,carnall2019vandels}.

Our focus here lies in the environmental component of galaxy evolution.  There is a large body of literature on the use of transition galaxies to trace quenching in dense environments
\citep[e.g.][]{zabludoff1996environment,balogh1999differential,poggianti1999star,poggianti2009environments,hogg2006triggers, yan2009deep,mok2013efficient, mok2014formation,paccagnella2017omegawings,paccagnella2019strong}.  This work was pioneered with the discovery of "E+A" galaxies in clusters \citep{DG83,dresser1992spectroscopy,CS87}; these are galaxies that have strong Balmer absorption lines in their spectra, indicative of recent star formation, but no signs of ongoing star formation since they lack [OII] or H$\alpha$ emission. 
Those E+A galaxies with the strongest Balmer absorption lines require a burst of star formation to precede a rapid quenching event \citep{CS87,2008ApJ...688..945Y,French18}, which has led to their classification as "post-starburst".  

The identification of traditional E+A galaxies galaxies requires spectroscopy.  An alternative approach has been to use the "green valley" in photometric data.  Originally the green valley was defined as the region between the red sequence and blue cloud in a colour-magnitude or colour-stellar mass diagram  \citep{schiminovich2007uvoptical,schawinski2014green,Vulcani15}.  However, this definition includes many dusty star-forming galaxies, especially at high masses and high redshifts.  This is mitigated by using rest-frame colour-colour diagrams like ($U-V$) vs ($V-J$), and again identifying the green valley as the region between the quiescent and star-forming populations  \citep{2011ApJ...736..110M,mok2013efficient,mok2014formation}.  

A further improvement is to use the rest-frame near-ultraviolet ($NUV$, $\lambda\approx 2300\AA$) colour when available, as this provides a greater separation between the star-forming and quiescent population and thus greater sensitivity to any transitioning galaxies 
\citep{salim2007uv,moutard2016vipers,moutard2018fast}. Others have used different combinations of colours \citep{wild2016evolution,leja2019beyond}, SFRs \citep{schiminovich2007uvoptical,salim2014green,Paccagnella2016} or spectral types \citep{2011ApJ...743..168K,2018ApJ...863..131F,2021arXiv210105820S} in order to identify transforming galaxies.

The evolution of environmentally-driven quenching is a promising avenue to explore further, as the dynamical time and growth rate of structure are largely decoupled from the cosmic star formation rate evolution at $z<2$.  
The Gemini Observations of Galaxies in Rich Early ENvironments (GOGREEN) survey \citep{balogh2017gemini,balogh2020gogreen} was undertaken for this reason, to study dense environments at $1<z<1.5$, with target clusters spanning a large range in halo mass representative of progenitors of today’s clusters.  
In previous work we have studied the star formation histories \citep{old2020gogreen,Webbgogreen}, morphologies \citep{Chanmorph} and SMFs \citep{Reevesgogreen,Changogreen,van2020gogreen} of the galaxies in these clusters.  A general picture is emerging in which environmental quenching plays a significant role in galaxy evolution even at these redshifts.  However, \citet{van2020gogreen} demonstrate that the SMF of quiescent galaxies is independent of environment, at least for galaxies $M>10^{10}~{\rm M}_{\odot}$.  One interpretation of this result is that many of these massive, quiescent cluster galaxies may have 
already been in place at much earlier times, rather than transformed following accretion onto an established cluster \citep[see also][]{Pogg06}.  In this case, the population of environmentally-driven transition galaxies would be small, or even absent, in these clusters.   

In this paper we make use of deep rest-NUV photometry of GOGREEN galaxies to select the intermediate green valley in colour-colour space at $1<z<1.5$.  This approach is taken to capture {\it all} galaxies as they pass from star-forming to quiescence, independent of the quenching mechanism.   It therefore provides a complete census of the transition rate, though it is not "pure" in the sense that it will also include substantial "contamination" from edge-on spiral galaxies and galaxies with slowly declining SFRs \citep[e.g.][]{schawinski2014green}. We therefore complement this with analysis of both "blue quiescent" (BQ) galaxies \citep{belli2019mosfire,carnall2019vandels} and spectroscopic post-starburst galaxies \citep[e.g.][]{Pogg06,muzzin2014phase},
to obtain a more complete picture of the recent quenching rate in these systems.

The structure of the paper is as follows. The data and galaxy selection are described in \S~\ref{sec-data}.  Our main results on the spatial distribution and SMF of green valley and other transition galaxies is presented in \S~\ref{sec-results}.   The implications of these results in the broader context of cluster galaxy evolution are discussed in \S~\ref{sec-discuss}.  Finally, we summarize our findings in \S~\ref{sec-conc}.
Throughout this paper, we adopt the AB magnitude system, a  \citet{chabrierIMF} initial mass function (IMF), and a flat $\Lambda$CDM cosmology with $H_0=70~{\rm km}~{\rm s}^{-1}~{\rm Mpc}^{-1}$, $\Omega_m=0.3$, and $\Omega_{\Lambda}=0.7$, unless otherwise specified.

\section{Data}\label{sec-data}
\subsection{The GOGREEN Survey}
The GOGREEN project is a spectroscopic and photometric survey of 21 galaxy clusters and groups at $1<z<1.5$.  For each system, deep ($\gtrsim 24$\,AB) imaging is available in $\sim 11$ bands between $u$ and 4.5$\mu$m.  The survey is built upon a large multi-object spectroscopic campaign with Gemini GMOS, targeting faint ($z' < 24.5$ and $[3.6 \mu m] < 22.5$) galaxies over a 5.5\arcmin\ field around each system.  The spectroscopic sampling is unbiased with respect to galaxy type for $M\gtrsim 10^{10.2}~\msun$, and secure redshifts are available for over $1500$ galaxies.  More details on the survey strategy and data are available in \citet{balogh2017gemini} and \citet{balogh2020gogreen}.

This work will focus on the eleven massive clusters\footnote{At the time of writing, $K$-band imaging observations were not complete for the final cluster, SpARCS1033+5753.} in the GOGREEN sample at $z<1.4$, selected  from the SpARCS \citep{wilson2009spectroscopic,muzzin2009spectroscopic,demarco2010spectroscopic} and SPT \citep{brodwin2010spt,foley2011discovery,stalder2013spt} surveys.  With one exception (SpARCS1034, which has $\sigma=250\pm30\,\mbox{km/s}$ based on only nine spectroscopic members), these systems have velocity dispersions of $500<\sigma/\,\mbox{km/s}\lesssim 900$
\citep{balogh2020gogreen}.  
The relevant cluster-specific parameters are provided in Table \ref{table-clusterinfo}. This includes the cluster radius $R_{200}$, defined as the radius within which the mean mass overdensity is 200 times the critical density at that redshift, and the corresponding mass $M_{200}$.  For most clusters, these are taken from \citet{BivianoGG}.  SpARCS0219 and SpARCS1034 have too few spectroscopic members for the analysis in that paper, and the estimates of their $R_{200}$ are obtained from the velocity dispersions in \citet{balogh2020gogreen}.

\begin{table}
\centering
\caption{Relevant properties of the eleven GOGREEN clusters analysed in this work.  These include the mean redshift (column 2), the virial radius $R_{200}$ (column 3) and mass $M_{200}$ (column 4), 
and the stellar mass limit of the photometric catalogues (column 5).}
	\label{table-clusterinfo}
\begin{tabular}{lllcl}
\hline
Cluster & Redshift & $R_{200}$  & $\log_{10}{\frac{M_{200}}{\mathrm{M}_{\odot}}}$  
& $\log_{10}{\frac{M_{\ast,lim}}{\mathrm{M}_{\odot}}}$ \\
  & &(Mpc) & 
  & 
  \\
  \hline

SPTCL-0205       & 1.320    &    0.76     & 14.3
&     9.9\\
SPTCL-0546       & 1.067    &    1.17     & 14.8
&     9.6\\
SPTCL-2106       & 1.131    &    1.23     & 14.9
&     9.8\\
SpARCS-0035      & 1.335    &    0.93     & 14.6
&     9.7\\
SpARCS-0219      & 1.325    &    0.79     & 14.4
&     9.9\\
SpARCS-0335      & 1.368    &    0.67     & 14.2
&     10.1\\
SpARCS-1034      & 1.385    &    0.24     & 13.0
&     9.5\\
SpARCS-1051      & 1.035    &    0.88     & 14.4
&     9.3\\
SpARCS-1616      & 1.156    &    0.92     & 14.5
&     9.6\\
SpARCS-1634      & 1.177    &    0.85     & 14.4
&     9.5\\
SpARCS-1638      & 1.196    &    0.71     & 14.2
&     9.5\\
\hline
\end{tabular}
\end{table}

\subsection{Galaxy Sample}\label{sec-sample}
For this paper we primarily use the photometric sample of galaxies, for which photometric redshifts, stellar masses and rest-frame colours have been obtained as described in \citet{van2020gogreen} and \citet{balogh2020gogreen}. 
In summary, the photometry was fit with stellar population synthesis models of \citet{BC03}, using the {\sc FAST} \citep{FAST} code, assuming star formation histories parameterised with a declining exponential function.  Rest-frame colours were computed from the best-fit models, in standard SDSS ($ugriz$), Johnson ($UBVRI$), 2MASS ($JHK$) and GALEX ($NUV$ and $FUV$) filters.  We note that the use of a smooth, parametric star formation history is known to underestimate the stellar mass by up to 0.3\,dex compared with nonparametric (binned) star formation histories \citep{2019ApJ...876....3L,Webbgogreen}. 

The photometric redshifts have been calibrated against the extensive spectroscopic sample to ensure they are unbiased over the redshift range of interest, $1<z<1.5$.  Small zeropoint corrections have been applied to the rest-frame colours to ensure they are consistent across all clusters in GOGREEN.  

Stellar mass completeness limits range from $9.5 \leq \log(M/\mathrm{M}_\odot) \leq 10.1$, and are given in Table~\ref{table-clusterinfo}.  These are computed by \citet{van2020gogreen}, based on the depth of the $K$-band photometry. For the purpose of this paper our sample is limited to 3062 galaxies with $\log(M/\mathrm{M}_\odot) \geq 9.5$, with good quality photometry in the $K$-band (SExtractor flags $<4$) and within the redshift range $1<z<1.5$.  We remove stars based on their colours, as described in \citet{van2020gogreen} and \citet{balogh2020gogreen}.

The spectroscopic subsample is representative of all galaxy types, for stellar masses $M\gtrsim 10^{10.2}~{\rm M}_\odot$ \citep{balogh2020gogreen}.  Of the 722 galaxies in our 11 clusters with secure redshifts $1<z<1.5$, 342 are above this mass limit.  In addition to the redshifts we will 
 make use of several relevant spectral indices.  The first is the strength of the 4000\,\AA\ break, D4000, using the definition of \citet{balogh1999differential}, which is relatively insensitive to the effects of dust.  We also measure the equivalent widths of H$\delta$ absorption and [O\, II] emission, again following the methodology of \citet{balogh1999differential}.  For [O, II], we 
identify the presence or absence of emission using the difference in the Bayesian Information Criterion, $\Delta_{\text{BIC}}$, which measures the extent to which model fits that include an emission line are preferred, as described  in \citet{old2020gogreen}.

\subsubsection{Cluster membership}\label{sec-membership}
For the photometric sample, candidate cluster members are selected using a photometric redshift ($ \pm 0.16$) and radius ($<1$\,Mpc) cut around the cluster centre.  The redshift cut is similar to what is used in \citet{van2020gogreen}, chosen as a suitable balance between including most genuine cluster members without too much contamination.  Nonetheless, since this redshift range is wide, driven by the uncertainties on those redshifts, a statistical background subtraction is required for the cluster population even after this selection.  Following \cite{van2013environmental}, we use the large spectroscopic subsample to derive a correction factor to the photometric sample in each radius or mass bin $i$, as:
\begin{equation}
    \text{corr}_i=\frac{N_{i\text{, secure cluster}}+N_{i\text{, false negative}}}{N_{i\text{, secure cluster}}+N_{i\text{, false positive}}},\label{eq-corrfactor}
\end{equation}
where $N_{i\text{, secure cluster}}$ are photometrically-selected and spectroscopically confirmed cluster members ($|\Delta z_{\rm spec}|<0.02$), $N_{i\text{, false negative}}$ are spectroscopically confirmed cluster members that lie outside the $\pm 0.16$ photometric redshift selection, and $N_{i\text{, false positive}}$ are galaxies that lie within the photometric redshift range but are spectroscopically confirmed to be foreground or background galaxies.  We calculate this correction factor for each galaxy subpopulation under consideration (described in the following subsections).
Throughout the paper we include all cluster galaxies, including the brightest cluster galaxy (BCG, which is sometimes omitted) in the analysis.
\begin{table*}
\centering
\caption{For each of the subpopulations considered in this paper, we present their definitions and sample size.  The final column provides the number of candidate cluster members.  For the photometric sample this includes all galaxies within $1$\,Mpc of the cluster centre and a redshift $|\Delta z_{\rm spec}|<0.16$.  For the spectroscopic PSB we report the number of dynamically identified cluster members.  
}
\label{table-photselections}
\begin{tabular}{llll}
\hline
{\bf Population} & {\bf Selection Criteria} & {\bf Total sample} & {\bf Cluster members} \\ \hline \hline
Star-forming (SF) & $ (NUV-V) < 2(V-J)+1.1 $& 1302&463\\ \hline
Quiescent (Q) & $ (NUV-V) > 2(V-J)+1.6 $& 702&504 \\ \hline
Green Valley (GV) & $ 2(V-J)+1.1 \leq (NUV-V) \leq 2(V-J)+1.6 $ & 257&125\\ 
\hline
\multirow{2}{*}{Blue Quiescent (BQ)} &
$ (V-J)+0.45 \leq (U-V) \leq (V-J)+1.35$\\
&$-1.25\,(V-J)+2.025 \leq (U-V) \leq -1.25\,(V-J)+2.7$ & 164&106
\\
\hline
Spectroscopic Post-starburst (PSB) & 
    $\left(\text{D4000} < 1.45\right) \cap \left(\Delta_{\text{BIC}}<-10\right)$ & 54&34  \\
\hline
\end{tabular}
\end{table*}

\begin{figure*}
    \centering
    \includegraphics[scale=0.55]{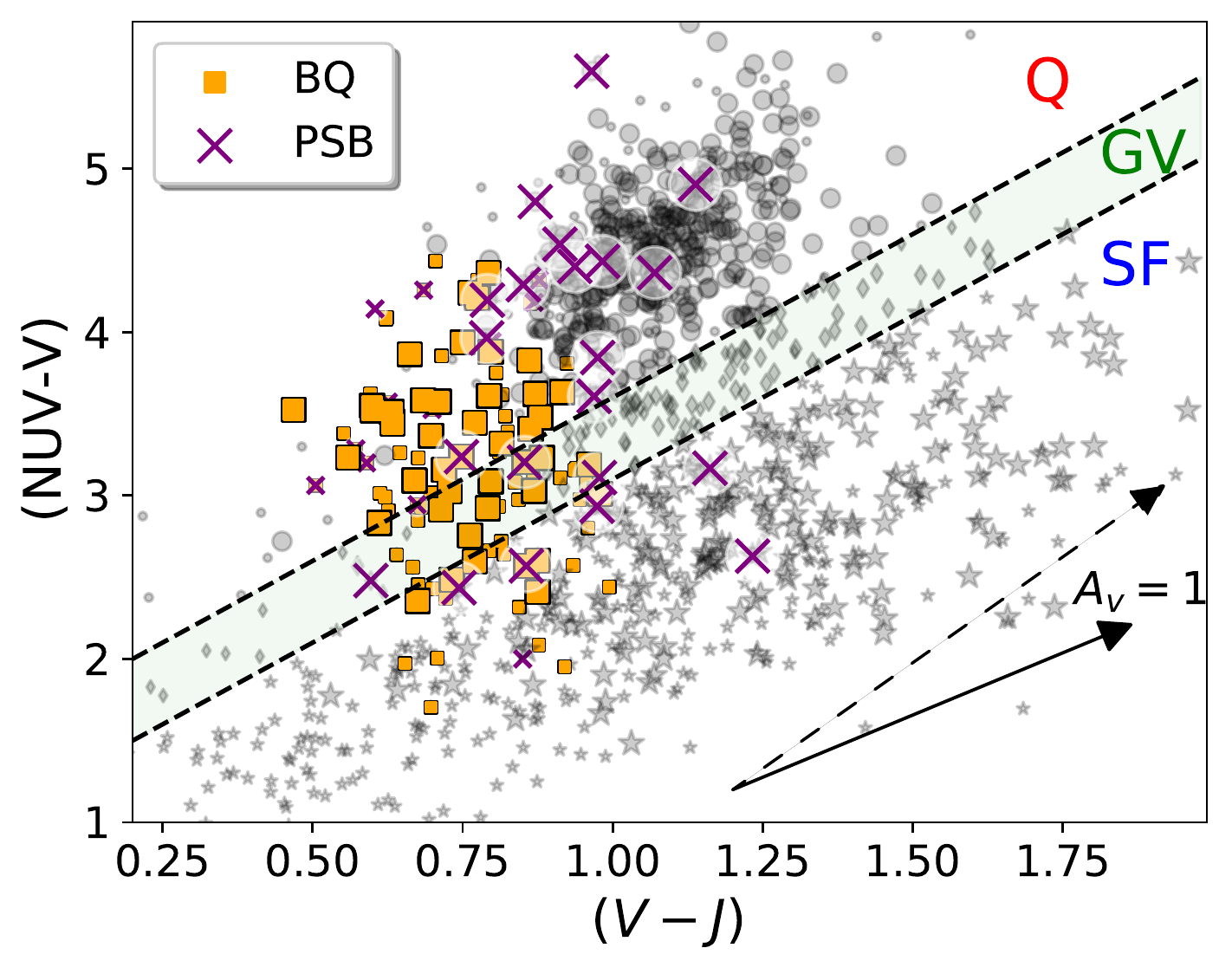}
    \includegraphics[scale=0.55]{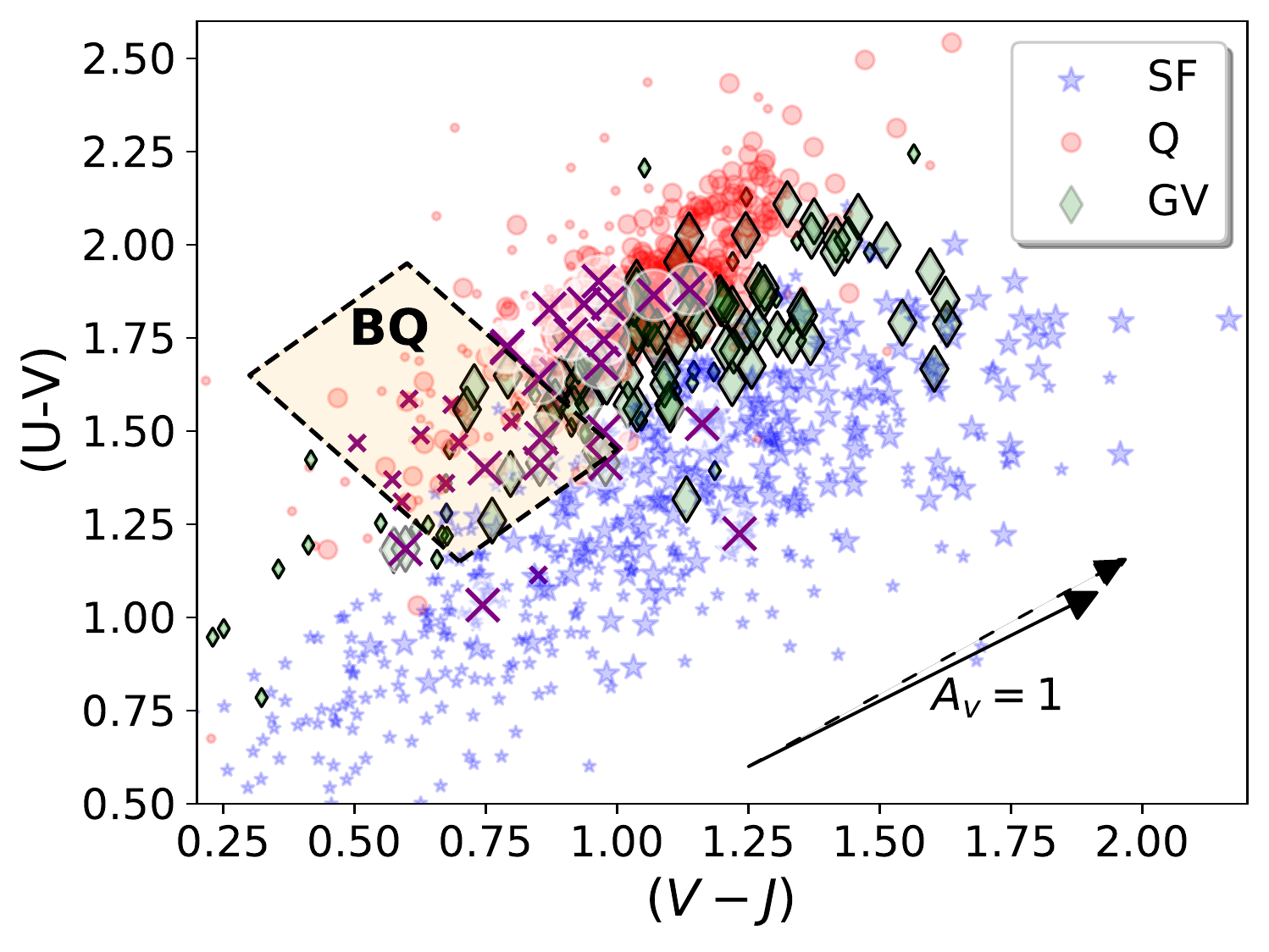}
    \caption{The {\it left panel} shows our cluster samples in the ${NUV}-VJ$ colour-colour plane that we use to define our primary Quiescent (Q), Star-forming (SF) and green valley (GV) populations.  
The BQ galaxies (orange squares) are selected in $(U-V)$ $(V-J)$ colour space, as identified by the orange filled box shown on the {\it right panel}.
The spectroscopic post-starbursts (purple crosses) are selected by their [O\,II] emission and D4000 break. 
Large symbols indicate galaxies above the spectroscopic mass limit of $M=10^{10.2}\,\mathrm{M}_\odot$.  We show extinction vectors for $A_V=1$\,mag, assuming a \citet[][solid line]{Calzetti}  or \citet[][dashed line]{CCM89} extinction law. } 
    \label{fig:photselect}
\end{figure*}

The comparison photometric field sample is selected from all galaxies with photometric redshifts $1<z<1.5$ and at $r>1.5$\,Mpc from the centre of each cluster. We do not excise galaxies at the cluster redshift, or spectroscopically confirmed members from this sample.  This may not be perfectly representative of a random field sample, and will include some galaxies associated with the cluster outskirts and surrounding structures\footnote{From the spectroscopic subsample, less than ten per cent of galaxies in this field sample with spectra are identified as cluster members by \citet{BivianoGG}.}. 
We elect not to use a larger, more representative field sample like UltraVISTA \citep{ultravista}; this ensures our definition of different populations, particularly the green valley, can be directly and fairly compared with the cluster sample in a straightforward way.  Our results are insensitive to the quantitative details of how our field is defined (in terms of the radial cut or redshift limits), and we show in \S~\ref{section- mass} that the stellar mass functions we measure for quiescent and star forming galaxies are in excellent agreement with those of \citet{van2020gogreen}.
Thus we do not expect our conclusions to be qualitatively sensitive to the definition of the reference field sample.  

When considering the spectroscopic subsample, we select cluster members as defined in \citet{BivianoGG}.  The field sample consists of all non-members with spectroscopic redshifts $1<z<1.5$.  We do not apply any incompleteness corrections to this sample, so integrated quantities should be treated with caution. Our main analysis considers the fraction of different galaxy types in bins of stellar mass, and this should be insensitive to such corrections above our spectroscopic completeness limit of $M=10^{10.2}~{\rm M}_\odot$.
 
\begin{figure}
    \centering
    \includegraphics[scale=0.45]{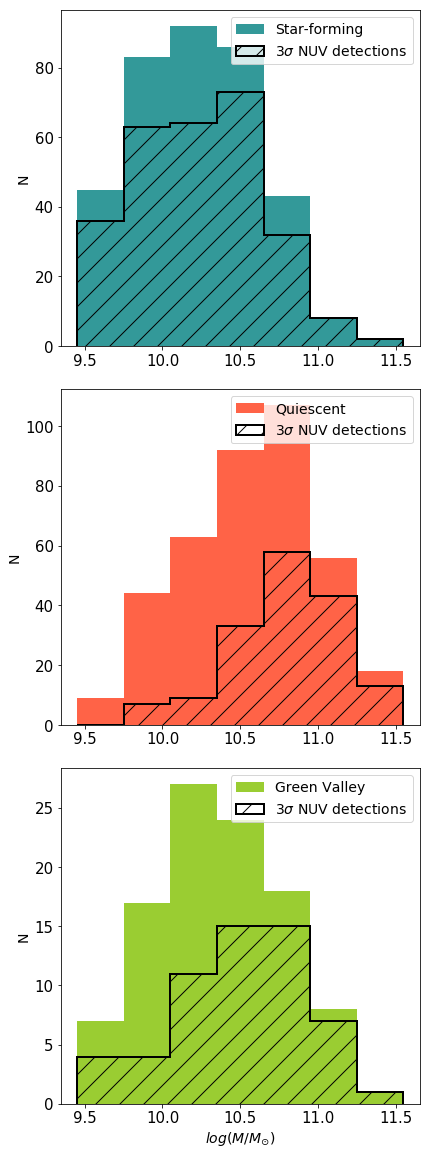}
    \caption{The fraction of galaxies that are detected (at $>3\sigma$) in either the $g$ or $V$ band filters, for each population SF, Q and GV.  These filters are close to the rest-frame $NUV$ at the redshifts of interest. }
    \label{fig:detect_fraction}
\end{figure}

\subsubsection{Star-forming, Quiescent and Green Valley galaxies}\label{sec-NUVJ}
We define three independent galaxy populations based on their positions in rest-frame ($NUV-V$) and ($V-J$) colour-colour space \citep{schawinski2014green,smethurst2015galaxy,moutard2018fast}, hereafter referred to as $NUV-VJ$, as shown in the left panel of Figure \ref{fig:photselect}. 
As has been discussed extensively in the literature, the galaxy distribution in this space shows two main populations.  The relatively tightly localized clump of galaxies with the largest $(NUV-V)$ colours are identified with quiescent (Q) galaxies, while the more extended sequence running from lower left to upper right are primarily star-forming (SF) galaxies.  We define an  intermediate region -- the green valley (GV) -- that is approximately parallel to the SF sequence, as shown in Figure~\ref{fig:photselect}.  The GV is  deliberately chosen to span the entirety of the colour-colour space, to ensure that any galaxy transitioning from star-forming to quiescence \textit{must} pass through it, regardless of the evolutionary path it takes. The 
width is arbitrarily defined, but selected to be a balance between making it small enough to reduce contamination from the other populations, while still wide enough to include a large population of transitioning galaxies.  The definitions of these three populations are summarized in Table \ref{table-photselections}.  The SF sequence, and our GV boundaries, are closely parallel to the extinction vector expected for a \citet{Calzetti} dust law, which should mitigate contamination of the GV from dusty SF galaxies.  The steeper slope associated with a \citet{CCM89} dust law could lead to more significant contamination at the upper end of the main sequence (but see \S~\ref{sec-compare}, below).  These dust vectors are both shown on Figure~\ref{fig:photselect}.  
 
The use of the $(NUV-V)$ colour diagnostic, providing more than five magnitudes of dynamic range between the bluest and reddest galaxies, is made possible by the very deep $V$ ($>25.3$) and $g$ ($>26$) band imaging of GOGREEN clusters \citep{van2020gogreen}, so that most galaxies above our mass limit are directly detected at wavelengths close to the rest-frame $NUV$.  In Figure~\ref{fig:detect_fraction} we show the fraction of each galaxy type that is detected at $>3\sigma$ in either the $g-$ band (northern clusters, with Subaru Suprimecam imaging) or $V-$ band (southern clusters, from VLT-VIMOS).  
For $M>10^{10.5}\,\mathrm{M}_\odot$, most SF and GV galaxies have rest $NUV$ detections.  Those SF galaxies that are undetected are typically red, $(NUV-V)>3.5$, and lie at the upper end of the SF sequence, where dust extinction is high.   Importantly, even at the lowest stellar masses we probe, about a third of the GV galaxies are directly detected.  Non-detections could, in principle, be redder in ($NUV-V$) than predicted, though this colour is still constrained by the fit to the redder filters in the spectral energy distribution (SED). 

\subsubsection{Blue Quiescent Galaxies}
We next consider another population defined photometrically, as galaxies at the blue end of the quiescent population identified in $UVJ$ colour-colour space.  This has been advocated by, for example, \cite{belli2019mosfire}, who use  deep rest-frame optical spectroscopy of 24 quiescent galaxies at $1.5 < z < 2.5$ to demonstrate that galaxies in this part of colour-colour space are distinctly younger than most of the quiescent population \citep[see also Figure E1 in ][]{Webbgogreen}, with  mass-weighted stellar ages of $300-800$ Myr.   \citet{belli2019mosfire} argue that galaxies passing through this space are a result of relatively fast quenching; assuming that dust disappears immediately after star formation stops, such galaxies will be very blue for a short time, until the remaining massive stars die out.  

While \citet{belli2019mosfire} and others refer to these as "post-starburst" galaxies, we prefer a more model-independent term, and will call them "Blue Quiescent" (BQ) Galaxies.  We choose to adopt the same BQ selection definition as \citet{belli2019mosfire}, including using the $UVJ$ diagram rather than the $NUV-VJ$ we use for the Q, SF and GV classifications in \S~\ref{sec-NUVJ}.  The boundaries of the BQ population are given in 
Table \ref{table-photselections} and shown on the right panel of Figure \ref{fig:photselect}.  

\subsubsection{Spectroscopic Post-starburst Galaxies}
\begin{figure}
    \centering
    \includegraphics[scale=0.55]{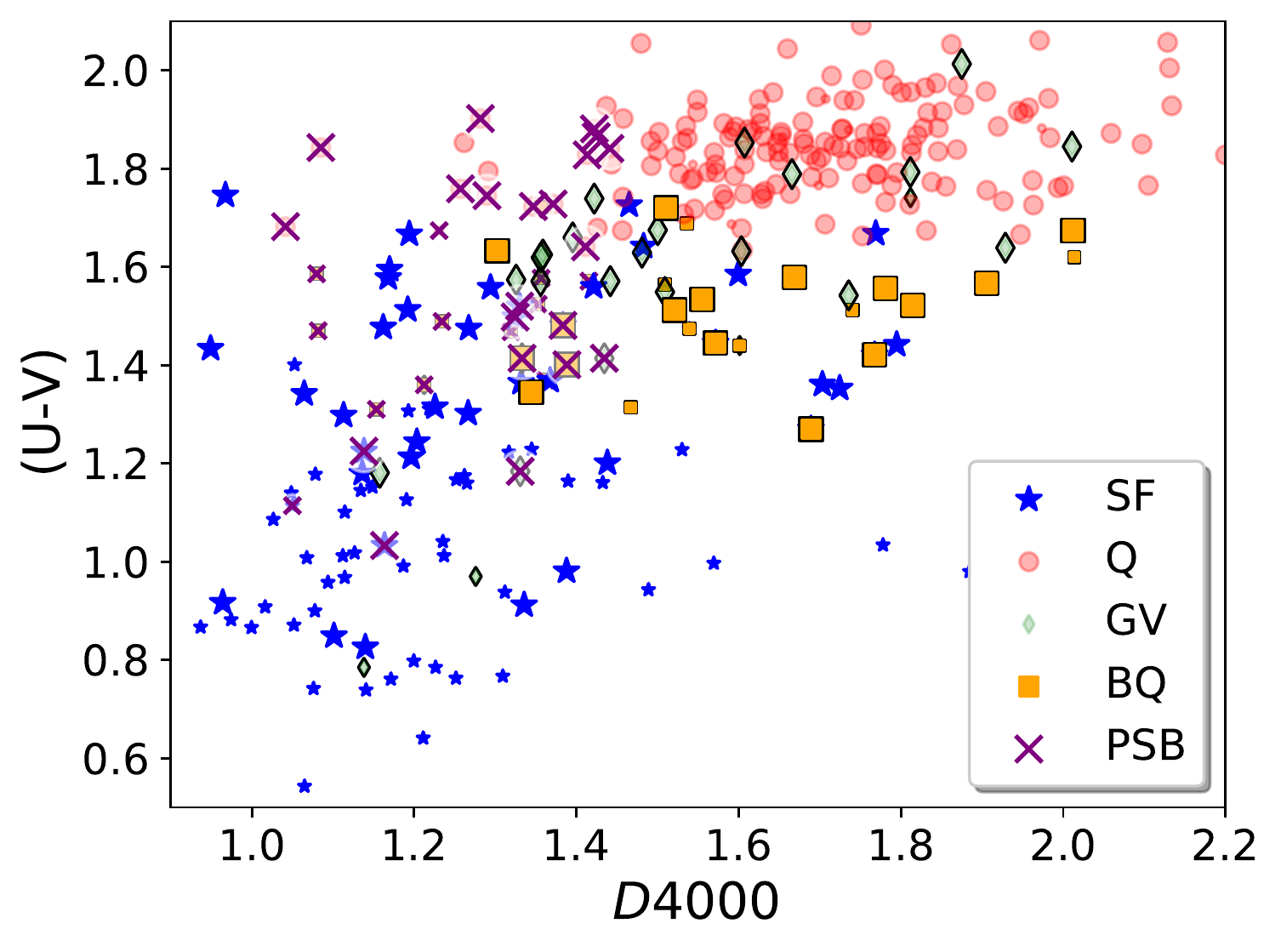}
    \caption{The $(U-V)$ colour of our cluster galaxies are shown as a function of the D4000 spectral index for spectroscopic cluster members.  Larger symbols indicate galaxies above the spectroscopic completeness limit of $M=10^{10.2}\,\mathrm{M}_\odot$.  There is a correlation, and our quiescent and star-forming galaxies are largely separable in both quantities.  The spectroscopic PSB sample is intermediate, and in particular includes several galaxies at the red boundary of the D4000 definition ($1.45$), with $(U-V)$ colours more typical of the Q population.  BQ and GV galaxies typically have larger D4000 indices, similar to Q galaxies, despite their bluer $(U-V)$ colours.}
    \label{fig-UV_D4000}
\end{figure}

The last population we consider are "post-starburst" (PSB) galaxies, based on the D4000 and [O\,II] spectral indices \citep{muzzin2014phase}, and thus limited to the spectroscopic sample. 
As Balmer absorption is difficult to measure reliably in low signal-to-noise ratio spectra, we follow \citet{muzzin2012} and use the 4000\,\AA\ break strength limit $\mathrm{D}4000<1.45$, rather than $H\delta$ strength, to identify young stellar populations. 
We identify the absence of [O\,II] emission using the difference in the Bayesian Information Criterion, $\Delta_{\text{BIC}}<-10$, where model fits without an emission line are strongly preferred, as described in \citet{old2020gogreen}.  This is different from \citet{muzzin2012}, who identify lack of emission by inspection of each spectrum.  We have tested the effects of varying the $\Delta_{\text{BIC}}$ threshold, or by adopting a threshold in rest-frame equivalent width, $W_\circ(\mbox{O\,II})<5$\,\AA.  Our results and conclusions are not sensitive to this choice.
Our final PSB definition is summarized in Table~\ref{table-photselections}.  We show where these galaxies are found in $UVJ$ and $NUV-VJ$ colour space as the purple crosses on Figure~\ref{fig:photselect}, and in the plane of $(U-V)$ and D4000 in Figure~\ref{fig-UV_D4000}. In general many of the spectroscopic PSB galaxies are quite red in $(NUV-V)$ or $(U-V)$ colours, clustered at the upper (red) end of the SF population in the $(U-V)$-D4000 plane.

\section{Results}\label{sec-results}
\subsection{Characteristics of the candidate transition populations}\label{sec-compare}
Figures~\ref{fig:photselect} and ~\ref{fig-UV_D4000} show where each of the candidate transition populations --- GV (green diamonds), BQ (orange squares) and PSB (purple crosses) --- lie in the different colour spaces used to define them.  These figures show that while there is some overlap between them,  they are largely independent samples.  The GV contains about a third of the BQ galaxies (and $22\pm 6$ per cent of GV galaxies are classified as BQ), but only $\sim 10$ per cent of the spectroscopic PSBs.  Those BQ and PSB galaxies that do lie in the GV are at the blue end of that region.

The GV galaxies, identified in $NUV-VJ$ space, are still largely found between the SF and Q galaxies in the $UVJ$ diagram (right panel of Figure~\ref{fig:photselect}).  This is notable because here the \citet{Calzetti} and \citet{CCM89} extinction vectors are nearly identical, and again parallel to the SF sequence.  If some of the GV galaxies identified in $NUV-VJ$ are due to dusty SF galaxies with a \citet{CCM89} extinction law, they would lie at the upper end of the SF sequence in the $UVJ$ diagram.  We do in fact find a few galaxies here, making up less than 5 per cent of the GV population.  

The PSB galaxies overlap substantially with the BQ population and the blue end of the GV in both colour-colour diagrams, but they also extend into the Q domain.  Interestingly, Figure~\ref{fig-UV_D4000} shows that both the BQ and GV galaxies typically have large D4000 break strengths characteristic of old populations, and larger than the threshold used to define PSB galaxies, despite their blue $(U-V)$ colours. 
\begin{figure}
    \centering
    \includegraphics[scale=0.52]{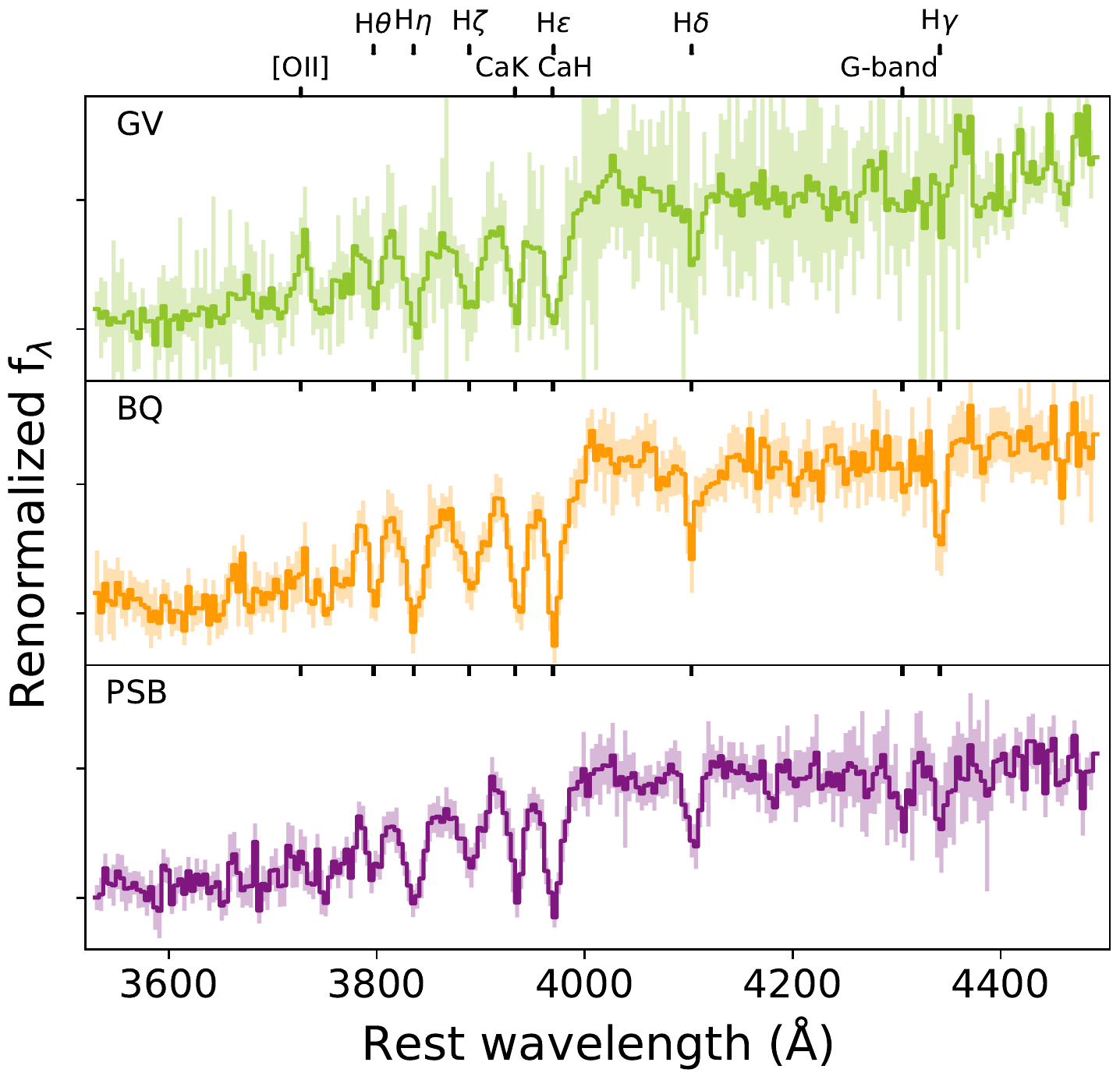}
    \caption{Median-combined spectra for all cluster galaxies in the GV, BQ or PSB categories considered in the text. Spectra are normalized just redward of the 4000\AA\ break and the shaded region represents the standard error on the median, from a bootstrap estimate. All three exhibit strong high-order Balmer and other absorption lines.  [O\,II] emission is weak but clearly present in both the BQ and GV samples.}
    \label{fig-specstacks}
\end{figure}
\begin{table}
    \centering
    \begin{tabular}{llll}
     {\bf Index} & {\bf D4000} & {\bf H$\delta$} & {\bf $[\mbox{OII}]$}\\ 
     &             & (\AA) & (\AA)\\
    \hline
    &\multicolumn{3}{c}{\bf Cluster}\\
    \hline
    {\bf GV}     & $1.5\pm 0.1$ & $4.1^{+4.2}_{-4.4}$ & $4.0^{+2.8}_{-2.6}$ \\
    \\
     {\bf BQ} & $1.52\pm 0.04$ & $4.8\pm 1.2$& $0.76\pm 1.6$\\
\\
    {\bf PSB}    & $1.34\pm 0.04$ & $3.7^{+1.3}_{-1.5}$  &$2.1^{+1.6}_{-1.7}$\\
    \hline
    &\multicolumn{3}{c}{\bf Field}\\
    \hline
     {\bf GV}     & $1.40\pm 0.05$ &  $3.1\pm 1.3$& $10.7^{+3.8}_{-4.1}$ \\
 \\
     {\bf BQ} & $1.39\pm 0.05$ & $4.6^{+1.3}_{-1.4}$& $7.4^{+3.0}_{-2.7}$\\
 \\
     {\bf PSB}    & $1.3\pm 0.1$ &$1.8^{+5.3}_{-5.1}$ &$2.1^{+5.0}_{-5.2}$\\
    \end{tabular}
    \caption{Spectral index measurements are shown for median combined spectra of each population type.  The $H\delta$ and [OII] indices are rest-frame equivalent widths, measured in \AA, defined such that they are positive in absorption for H$\delta$, and positive in emission for [OII]. Uncertainties are 68 per cent confidence limits obtained by Monte-Carlo resampling the measurements assuming the uncertainties on each spectral pixel are Gaussian distributed.}
    \label{tab:indices}
\end{table}

\begin{figure*}
\includegraphics[scale=0.55]{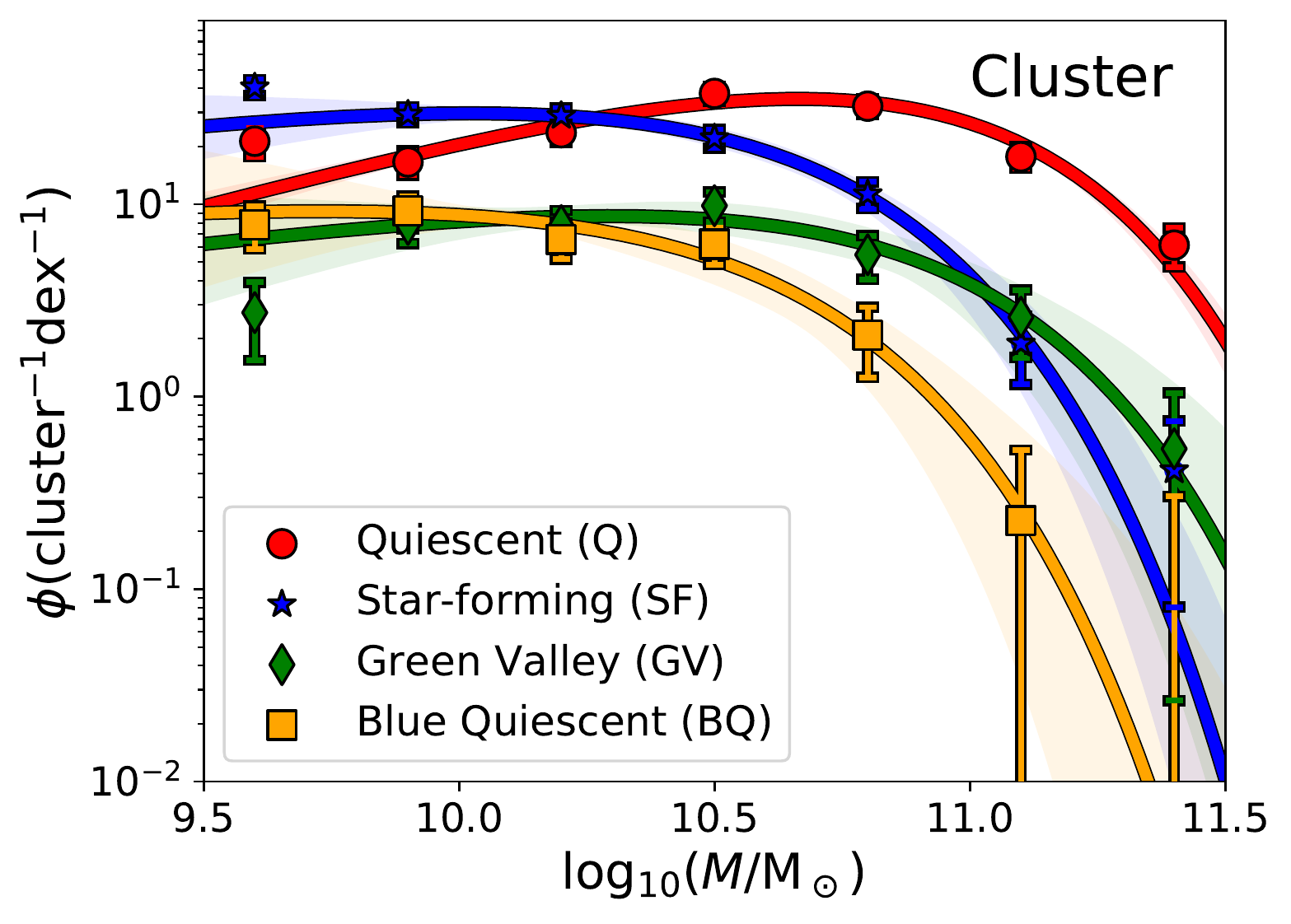}
\includegraphics[scale=0.55]{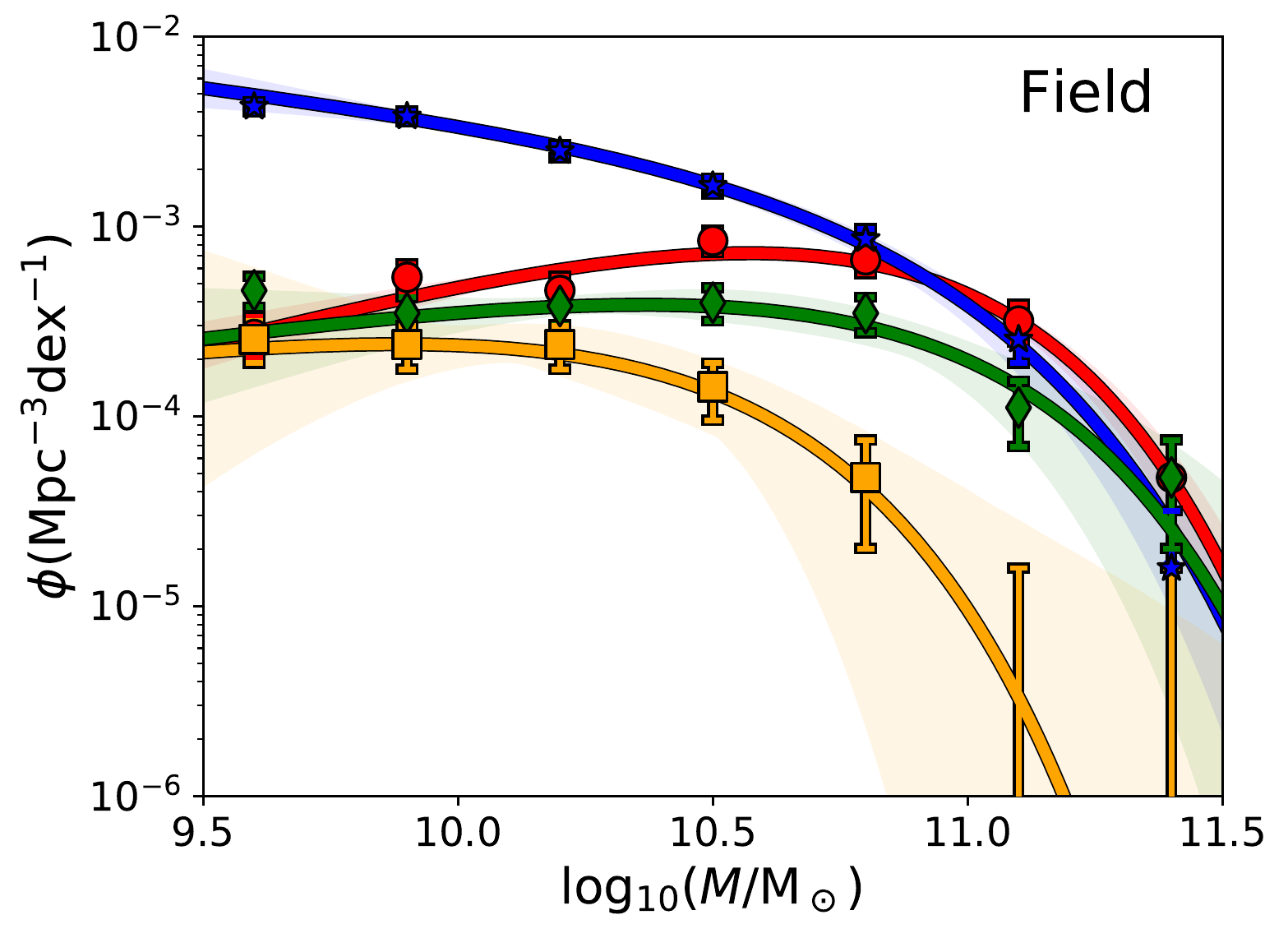}
\caption{The stellar mass functions for the star forming (blue), quiescent (red), green valley (green) and blue quiescent (orange) populations are shown for the cluster (left panel) and the field (right panel).   Schechter functions fits are shown as smooth curves, with shaded regions reflecting the 68 per cent confidence range. 
    Note that the BQ population is not independent of the others, and is defined in $UVJ$ space while the other three populations are defined in $NUV-VJ$ space.  The parameters of the best-fit Schechter function are given in Table \ref{table:massfits}.  The lowest mass bin is excluded from these fits, as described in the text.}
    \label{fig:gvmassfuncts}
\end{figure*}
We compare the average of the available spectra for cluster members of each galaxy type, in Figure~\ref{fig-specstacks}.  The spectra are normalized to the wavelength range $4115<\lambda/$\AA~$<4125$ and then median-combined. 
For each combined spectrum we measure the D4000 index, and the rest-frame equivalent widths of [OII] emission and H$\delta$ absorption as defined in \citet{balogh1999differential}.  The results are shown in Table~\ref{tab:indices}, both for the cluster sample shown in Figure~\ref{fig-specstacks}, and for the corresponding field populations.  
All selections show relatively strong Balmer absorption lines characteristic of young stellar populations, though the uncertainties are large.  
As reflected in Figure~\ref{fig-UV_D4000}, the GV and BQ galaxies exhibit significantly stronger D4000 than the PSB galaxies.  The average cluster GV spectrum shows weak [O\,II] emission, which could indicate residual star formation but could also result from a somewhat heterogeneous population that includes some normal star-forming galaxies. The [OII] emission is stronger and more significant in the field GV and BQ samples. 

Figures~\ref{fig:photselect}, ~\ref{fig-UV_D4000} and ~\ref{fig-specstacks} do not clearly identify a strict evolutionary sequence between the GV, PSB and BQ galaxies.  The colour evolution of a galaxy is sensitive not only to its star formation rate evolution during quenching, but to assumptions about the dust distribution and its evolution, and its prior star formation history \citep[e.g.][]{belli2019mosfire,French18}. It may be that these populations partly reflect different quenching "pathways".  While it would be informative to use stellar population modeling to better understand the relationships between these galaxies, we will leave that for future work and in the present paper focus on the demographics to provide insight into the dominant timescales associated with environmental quenching.

\subsection{Stellar Mass Functions}
\label{section- mass}
The SMFs for the Q, SF, GV and BQ populations in the cluster and field populations are shown in Figure \ref{fig:gvmassfuncts}. Note that the Q, SF and GV galaxies are independent, and together sum to the total cluster population.  The BQ galaxies, however, are not independent of these classifications.  Uncertainties assume Poisson statistics and include an additional uncertainty associated with the membership correction factor in Equation ~\ref{eq-corrfactor}.  We apply corrections for incompleteness in the $K-$band catalogue and for the different stellar mass limits of each contributing cluster, as described by \citet{van2020gogreen}.  The cluster mass functions are normalized to represent the number of galaxies per dex per cluster, while the field mass functions are normalized by the volume of the survey between $1<z<1.5$.  
We do not include the PSB population here, because they are drawn from the spectroscopic sample where large and uncertain completeness corrections are required; we will consider their relative abundance subsequently.   Finally, as stated in \S~\ref{sec-membership}, we do not exclude BCGs from this analysis.

The binned data are fit with Schechter functions of the form
\begin{equation}
    \label{schect}
    \phi(M)dM=\frac{dN}{dM}dM=
    \phi^*\left(\frac{M}{M^*}\right)^{\alpha}\exp\left(\frac{-M}{M^*}\right)\,dM.    
\end{equation}
or, for $x=\log_{10}{M/\mathrm{M}_\odot}$ as plotted in the Figure,
\begin{align}
    \label{schect_ln}
    \phi(x)dx&=\frac{dN}{dx}dx=
    \ln{10}\left(\frac{dN}{dM}\right)Mdx=\nonumber\\
    &=\ln{10}\,\phi^*M^*\left(\frac{M}{M^*}\right)^{1+\alpha}\exp\left(\frac{-M}{M^*}\right)\,dx.    
\end{align}
Following\footnote{\citet{van2020gogreen} fit the data for the SF population to a lower stellar mass limit, of $M>10^{9.5}\,\mathrm{M}_\odot$.  As the fit parameters are sensitive to the mass limit chosen, we elect to fit all our populations over the same, conservative, mass range.} \citet{van2020gogreen} we only fit the data with $M>10^{9.7}\,\mathrm{M}_\odot$, as the completeness corrections are both large and systematically uncertain (due in part to sparse and biased spectroscopic sampling) below that limit.  
The normalization parameter is fixed by forcing the integral to match the sum of the data where $\log(M/\mathrm{M}_\odot) \geq 9.7$; the other two parameters are fit by minimizing the $\chi^2$ statistic.  There are six data bins, omitting the lowest mass point, and two parameters, for $\nu=4$ degrees of freedom.   

\begin{figure*}
    \centering
    \includegraphics[width=8.5cm]{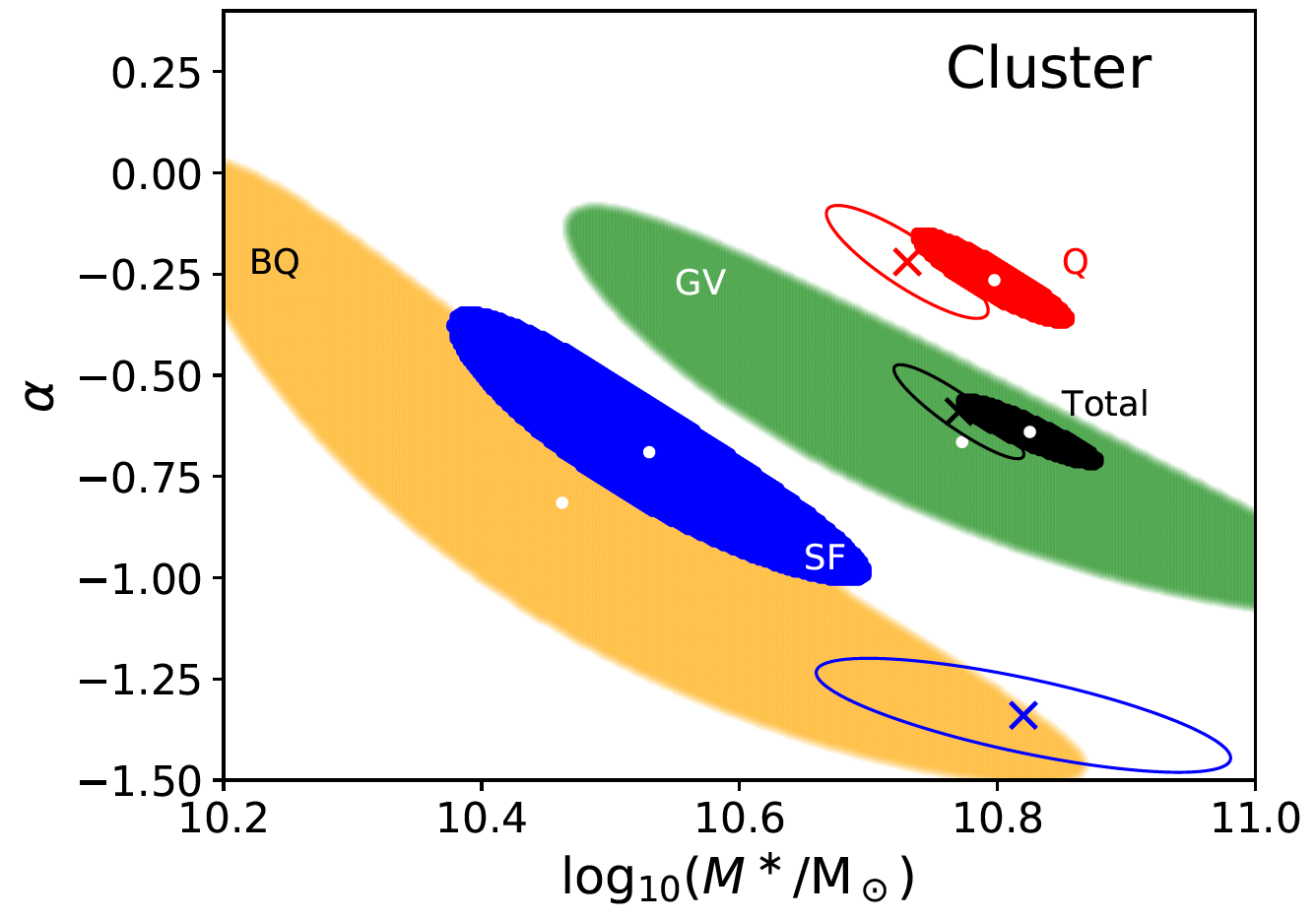} 
    \includegraphics[width=8.5cm]{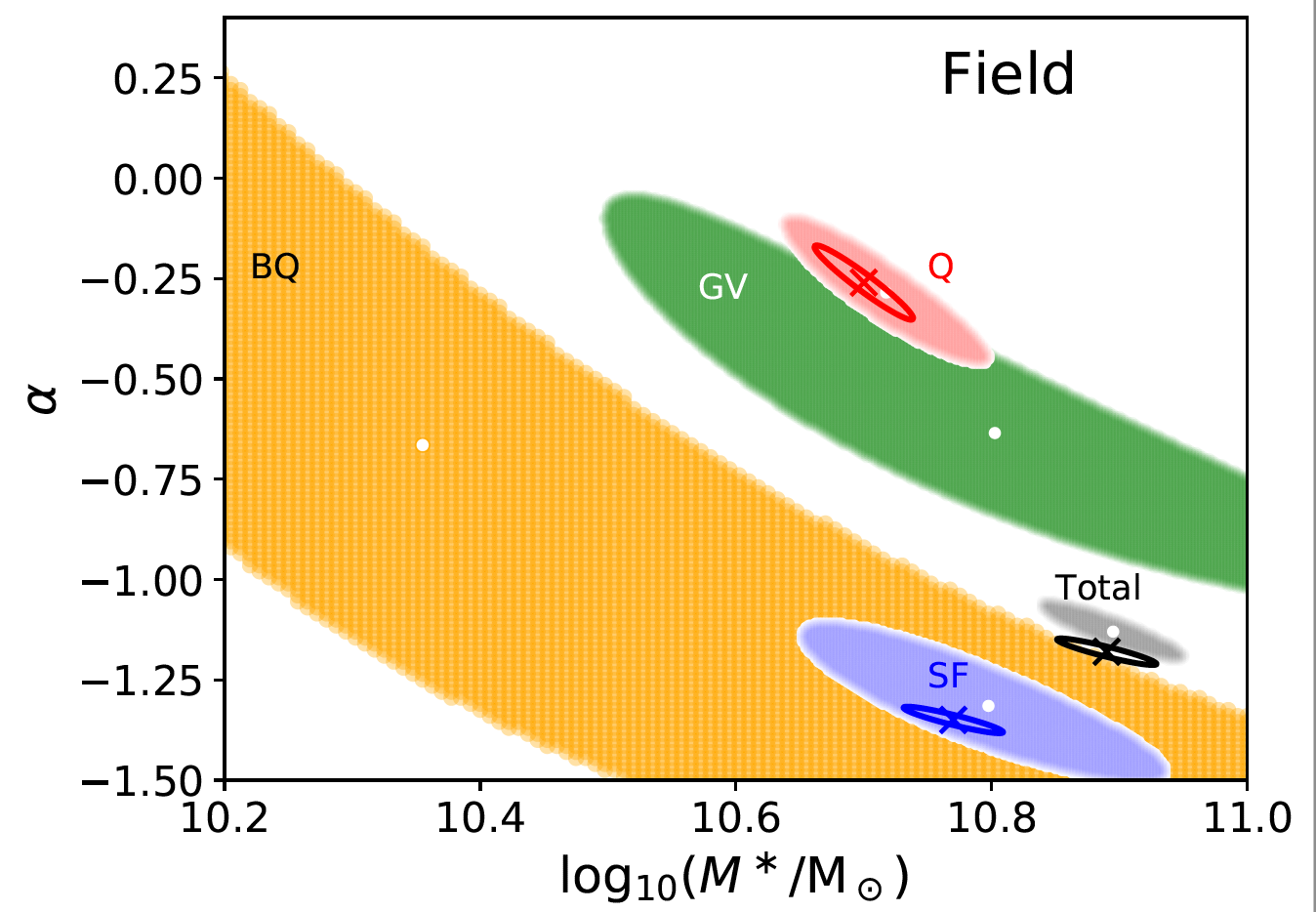} 
    \caption{We show the 68 per cent confidence limits on the $M^\ast$ and $\alpha$ parameters of our Schechter function fits, for the total (black) quiescent (red), star-forming (blue), green-valley (green) and blue quiescent (orange) galaxies.  The white dots represent the best fit.  Parameters for the cluster population are shown in the {\it Left} panel, and parameters for the field population are on the {\it Right} panel.  The crosses are the best fit parameters to the total, quiescent and star-forming populations (defined in $UVJ$ space, without a green valley) by \citet{van2020gogreen}; open ellipses indicate their 1$\sigma$ confidence ranges.   The parameters of the BQ population are not well constrained and the 68 per cent confidence limit extends beyond the limits of this figure. }
    \label{fig:mstaralpha}
\end{figure*}
The resulting fit parameters are given in Table~\ref{table:massfits} and we show the 68 per cent confidence ranges\footnote{For cases where the best fit has $\chi^2/\nu>1$, we temporarily increase the error bars to achieve $\chi^2=1$ and then find the 68 per cent confidence range.} of $M^\ast$ and $\alpha$ in Figure~\ref{fig:mstaralpha}.  These are compared with the SMF fit parameters of \citet{van2020gogreen}, which were based on a different classification, using $UVJ$ colours, and only separated galaxies into two types: quiescent and star-forming.    
For the total cluster sample, as well as the Q population, the fit parameters we derive are fully consistent with those of \citet{van2020gogreen}.  For the SF galaxies we find a somewhat\footnote{This difference is due primarily to the fact that we only fit the data above $M>10^{9.7}\,\mathrm{M}_\odot$, while \citet{van2020gogreen} include data above $M>10^{9.5}\,\mathrm{M}_\odot$ when fitting the SF population.} larger $\alpha$ and smaller $M^\ast$.  For the field sample, the fit parameters for the total, Q and SF populations are in excellent agreement with \citet{van2020gogreen}, who used the independent, larger UltraVISTA \citep{ultravista} survey.

\begin{table}
\caption{The values for the two-parameter Schechter mass function fits, for each of the galaxy populations in both the cluster and the field. The values in this table correspond to the parameters in Equation \ref{schect}, and are used to plot the SMF fits shown in Figure \ref{fig:gvmassfuncts}. For the cluster, the units of $\phi^\ast$ are $10^{-10}{\rm cluster}^{-1}{\rm dex}^{-1}$, while for the field the units are $10^{-15}{\rm Mpc}^{-3}{\rm dex}^{-1}$.  The $\chi^2$ values are calculated for six data points (the lowest mass bin is excluded from the fit) and two parameters (the normalization is fixed), yielding 4 degrees of freedom. }\label{table:massfits}
\begin{tabular}{llllll}
\hline
 &       & \textbf{$\chi ^2$} & \textbf{$\log_{10}(M^\ast/\mathrm{M}_\odot)$} & \textbf{$\alpha$} & \textbf{$\phi^\ast$} \\ \hline
{Cluster }    & SF    & 1.1              & $10.53$                      & $-0.69$             & 7.0              \\
              & Q   & 5.4              & $10.80$                      & $-0.26$            & 6.4             \\
            & GV    & 1.4               & $10.77$                      & $-0.66$             & 1.3           \\
           & Total & 6.0                & $10.82$                   & $-0.64$             & 8.8            \\ \hline

\multirow{3}{*}{Field}                & SF    & 1.6              & $10.80$   & $-1.31$            & 13.4             \\
                                      & Q   & 5.3             & $10.72$                      & $-0.28$   & 15.6             \\
                                      & GV    & 1.3               & $10.80$      & $-0.63$             & 5.5             \\
                                      & Total & 10.7              & $10.89$         & $-1.13$             & 19.5          \\
\hline
\end{tabular}
\end{table} 
In both the cluster and field environments, the shape of the GV SMF is intermediate between that of the SF and Q galaxies. Though the parameters are not tightly constrained, they are inconsistent with the SF population fit.    It is therefore unlikely that the majority of this intermediate population can be made up of galaxies transformed directly from the infalling SF population, without a significant change in stellar mass.   
On the other hand, the SMF of cluster BQ galaxies is fully consistent with that of the SF galaxies  (Figure~\ref{fig:mstaralpha}).  Though the fit is not well constrained, particularly in the field, due to the small sample size, the fit parameters are highly inconsistent with those of the Q population.  
This, together with their rarity in the field population discussed below, supports an interpretation in which most of the cluster BQ galaxies arise from rapid quenching of the SF galaxies after infall into the cluster environment.   

Throughout the rest of this paper we represent the SMF of each population as $\phi_i^j$, where $i\in\{SF, GV, Q, BQ\}$ represents the subpopulation type, and $j\in\{C,F\}$ identifies the environment as cluster or field.   The total SMF in each environment is just written as $\phi^C$ or $\phi^F$, with no subscript. 
Normalizing the SMF for each of the populations by the total SMF in the relevant environment converts the $\phi_i^j$ into fractions $f_i^j$, which allows us to compare the cluster and field distributions independently of any overall difference in SMF shape.

\subsection{Population abundance distributions}
\begin{figure*}
    \centering
    \includegraphics[scale=0.55]{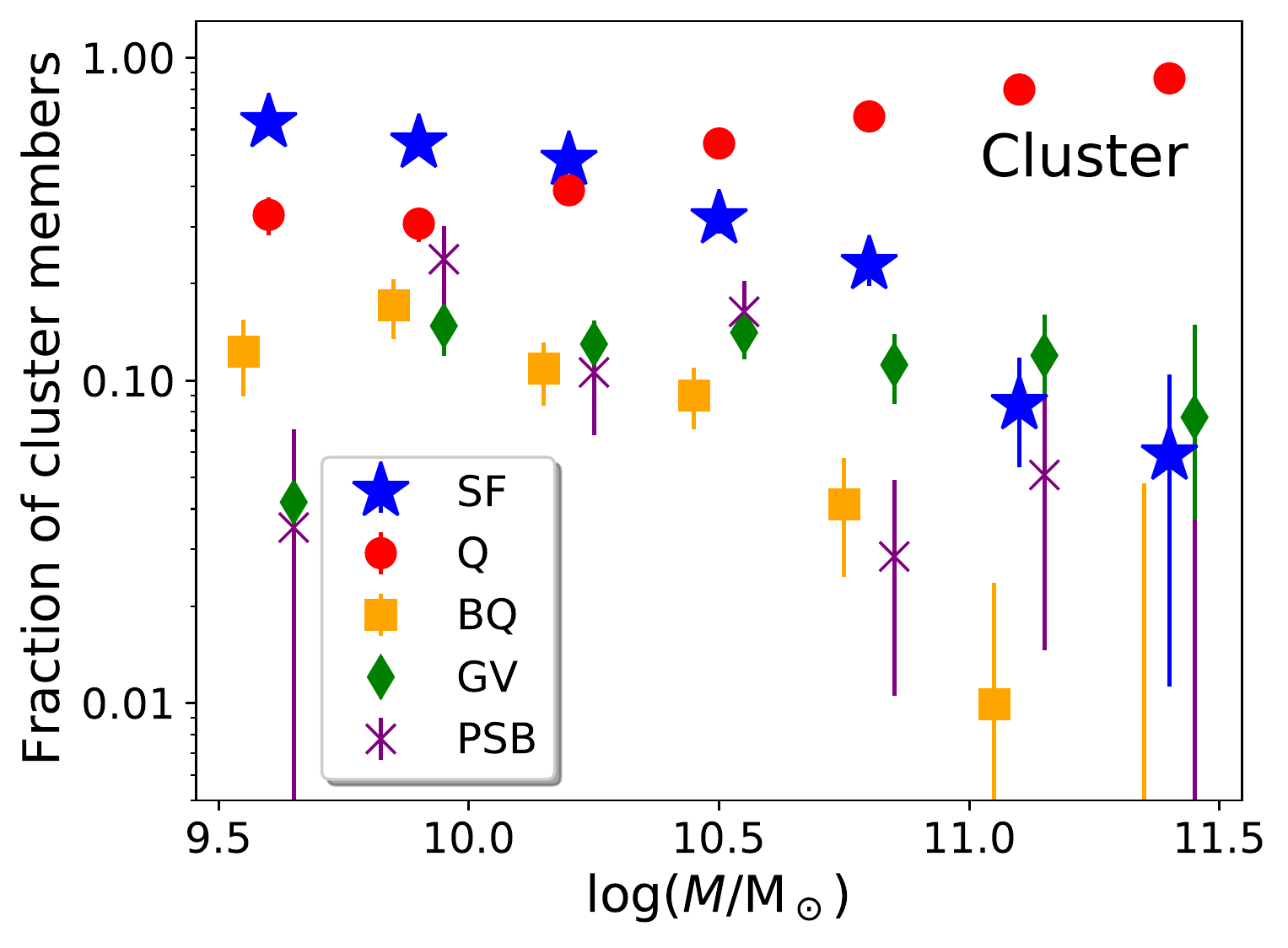}
    \includegraphics[scale=0.55]{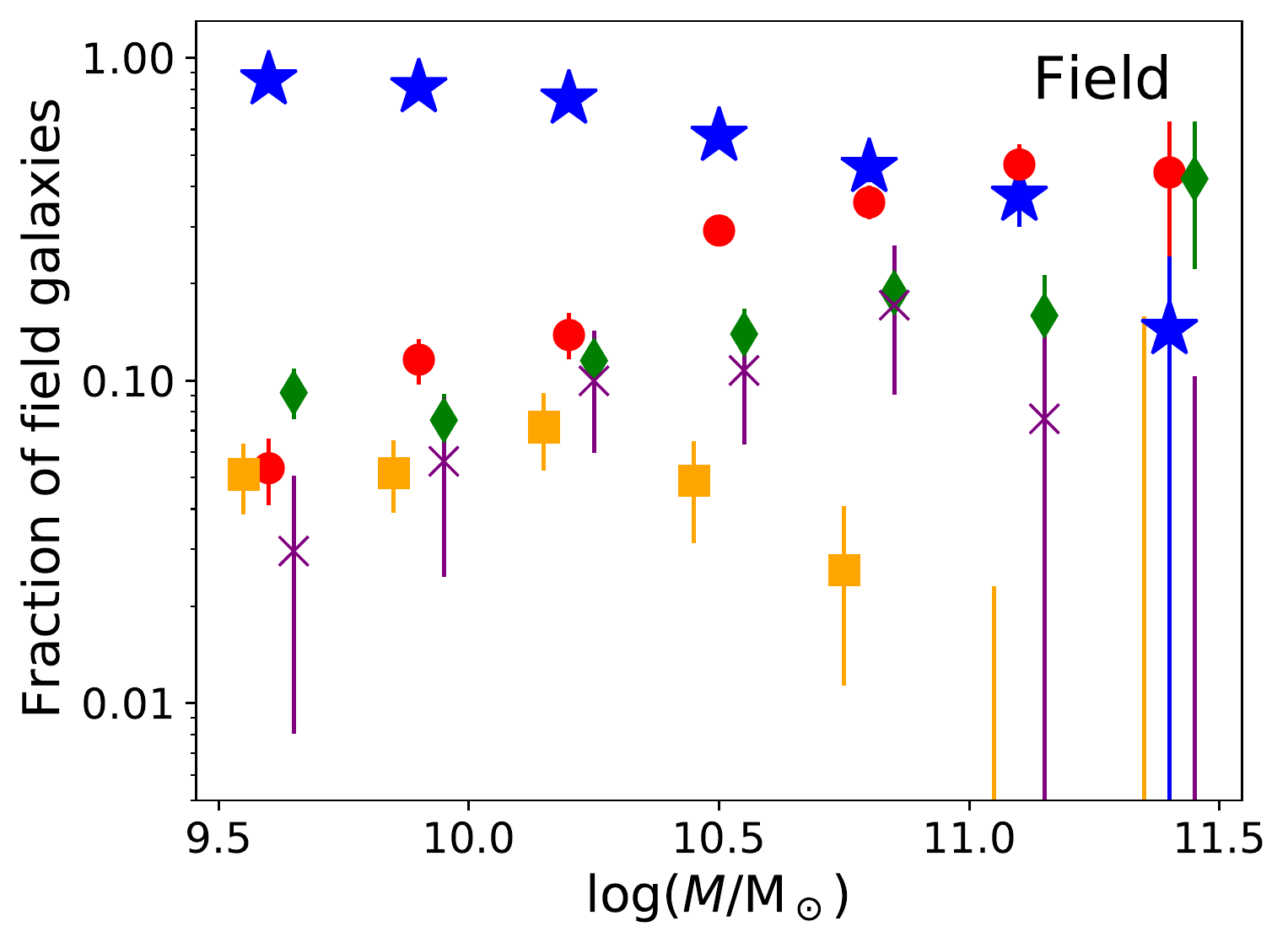}
    \caption{{\it Left: } The abundance of each cluster population, relative to the total, is shown as a function of stellar mass.  For the SF, Q, BQ and GV galaxies, this is measured from the SMFs presented in Figure~\ref{fig:gvmassfuncts}.  For the spectroscopic PSB galaxies we show the fraction of spectroscopically confirmed PSB cluster members relative to the total sample of spectroscopic cluster members; these values should be treated with caution for $M<10^{10.2}{\rm M}_\odot$, as the spectroscopic sample is not unbiased at those masses. The GV and PSB populations are not independent of the other three, so the fractions do not sum to unity in a given bin.  {\it Right:} The same, but for the field population.  For the photometric sample, this is all galaxies $>1.5$\,Mpc from the cluster centre, in the redshift range $1<z<1.5$.  For the PSB population we take all spectroscopic non-cluster members, also at  $1<z<1.5$, as determined by \citet{BivianoGG}.  Small offsets in the plotted $x-$position for the BQ and GV points are applied for clarity.
    }
    \label{fig:SMF_fracs}
\end{figure*}

\subsubsection{Stellar mass dependence}
In Figure~\ref{fig:SMF_fracs} we show the abundance of each subpopulation as a fraction of the total, in bins of stellar mass, for the cluster and field environments.   Uncertainties, here and elsewhere in the paper, are obtained from 1000 Monte Carlo samples of our galaxy populations and represent the 68 per cent confidence intervals. We include the spectroscopic PSB sample here (recall that the BQ and PSB galaxies are not independent of the other three classifications).  As the spectroscopy is unbiased with respect to type for galaxies $M\gtrsim10^{10.2}\,\mathrm{M}_\odot$ \citep{balogh2020gogreen}, the relative PSB abundance in bins of stellar mass is robust.  Below this mass limit the spectroscopic sample is incomplete, and the ratios should be used with caution.  The photometric sample of SF, Q, GV and BQ galaxies are representative down to $M\approx 10^{9.5}\,\mathrm{M}_\odot$, though we caution that the lowest mass bin has large and uncertain completeness corrections (and hence was excluded from the SMF fits).

The BQ galaxies exhibit a strong correlation with stellar mass, thier relative abundance increasing toward lower masses.  This closely parallels the trend seen in the SF galaxies, as expected given the similarity in their SMF shapes.  The cluster PSB sample also shows a similar behaviour, though the uncertainties are considerably larger.  This suggests that their SMF is also similar in shape to that of the BQ and SF galaxies, and is  consistent with a scenario in which they arise from approximately mass-independent quenching of the infalling SF galaxies.  The GV galaxies, on the other hand, show a more modest trend with mass in the cluster environment.  
All three cluster transition populations show a drop in abundance in the lowest stellar mass bin, but this should be treated with caution given the relatively large and uncertain incompleteness and cluster membership corrections\footnote{Specifically, at these masses the SF galaxies are the dominant population, and the contamination from field galaxies in the cluster sample is high, leading to a large correction.  An overestimate of the cluster SF population in this bin would lead to underestimates of the GV, BQ and PSB fractions relative to the total.} at this mass. In the field, both the GV and PSB populations show a decreasing fraction with decreasing mass, opposite to what is observed in the cluster environment, and suggesting that they may have a physically distinct origin.  We note here also that the abundance of massive ($M>10^{10.8}\mathrm{M}_\odot$) BQ galaxies in the field, $(1.4\pm 1)\times 10^{-5}$ Mpc$^{-3}$, is in  agreement with the $(0.9\pm 0.1)\times 10^{-5}$ measured by \citet{belli2019mosfire} at the same redshift, and corresponds to $1.8$ per cent of the total.

\begin{figure}
    \centering
    \includegraphics[scale=0.58]{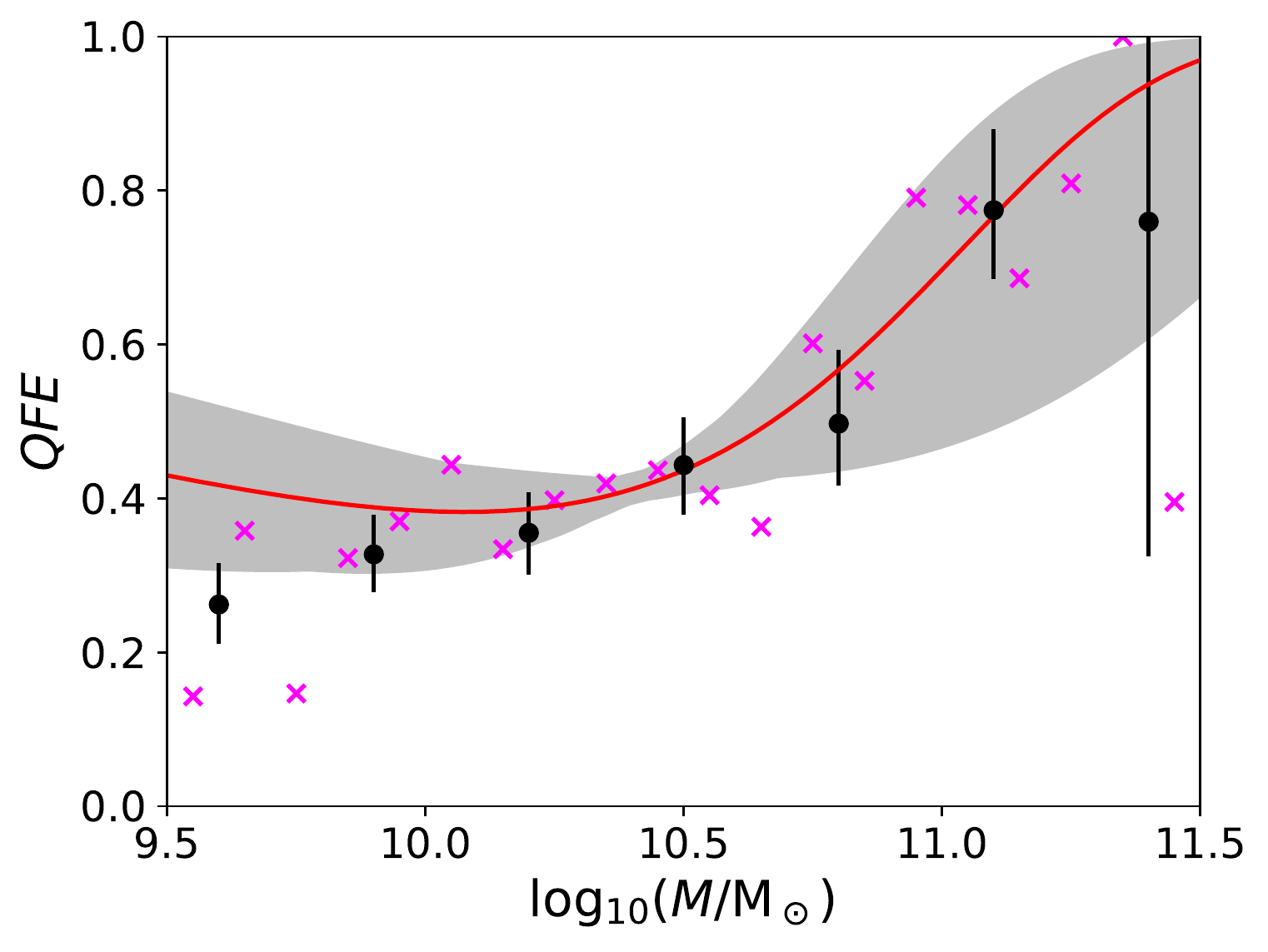} 
    \caption{The QFE (Equation~\ref{eq-QFE}) is shown as a function of stellar mass, based on the measured SMFs of the star-forming galaxies in the cluster and field sample.  Black points represent the binned data as medians with 68 per cent confidence limit error bars.  The measurement inferred from the Schechter function fits is shown as the red line for the best-fit parameters, and the grey shaded region for the 68 per cent confidence region.  
    The magenta crosses are the QFE measurements from \citet{van2020gogreen}, using the same data but a standard UVJ classification of star-forming galaxies.  We omit error bars on these points for clarity; they are somewhat larger than the error bars on our measurements because of the smaller bin sizes.}
    \label{fig:QFE_new}
\end{figure}
The most significant, and well-known, difference in cluster galaxy populations is the excess abundance of Q galaxies, and corresponding lack of SF galaxies, as seen in Figure~\ref{fig:SMF_fracs}.  This can be characterized by a "Quenched Fraction Excess"\footnote{We adopt this terminology as the most accurate description of the quantity, following \citet{wetzel2012galaxy, BaheHydrangea2017, van2020gogreen,Reevesgogreen}. Similar definitions in the literature refer to this quantity as the ``transition fraction'' \citep{vandenBosh2008satQuenching}, ``conversion fraction'' \citep{Balogh16, Fossati2017}, and ``environmental quenching efficiency'' \citep{Peng2010, Wetzel2015localSatDwarfQuenching, Nantais2017, vanderBurg2018}. } \citep[QFE; e.g.][]{vandenBosh2008satQuenching,Peng2010,BaheHydrangea2017,van2020gogreen} given by:
\begin{equation}\label{eq-QFE}
\mbox{QFE}\equiv\frac{f_{\rm SF}^F-f_{\rm SF}^C}{f_{\rm SF}^F},
\end{equation}
which represents the fraction of field SF galaxies at the epoch of observation that need to be quenched to represent their abundance in clusters at the same time.  By comparing the field and cluster at the same redshift, this measures the quenching that needs to occur for reasons that are correlated with environment, over and above the "natural" evolution of field galaxies.
We show the QFE as a function of stellar mass in Figure~\ref{fig:QFE_new}, both using the binned data and the smooth Schechter function fits to the SMFs.  We also compare with the analysis of \citet{van2020gogreen}, who used the same cluster data but a different galaxy classification and an independent field sample taken from UltraVISTA \citep{ultravista}.  Despite these differences, our measurements are in good agreement with theirs.  As observed by others \citep[e.g.][]{Balogh16,2017ApJ...847..134K,van2020gogreen}, the QFE increases steadily with stellar mass, though we note that over the substantial mass range $10^{9.75}<M/\,\mathrm{M}_\odot<10^{11}$, the data are consistent with a mass-independent $QFE\sim 0.4$. It is also remarkable that this value of $0.4$ agrees very well with the QFE measured at low redshift for all satellite galaxies \citep{peng2012}.   
 
\begin{figure}
    \centering
    \includegraphics[scale=0.53]{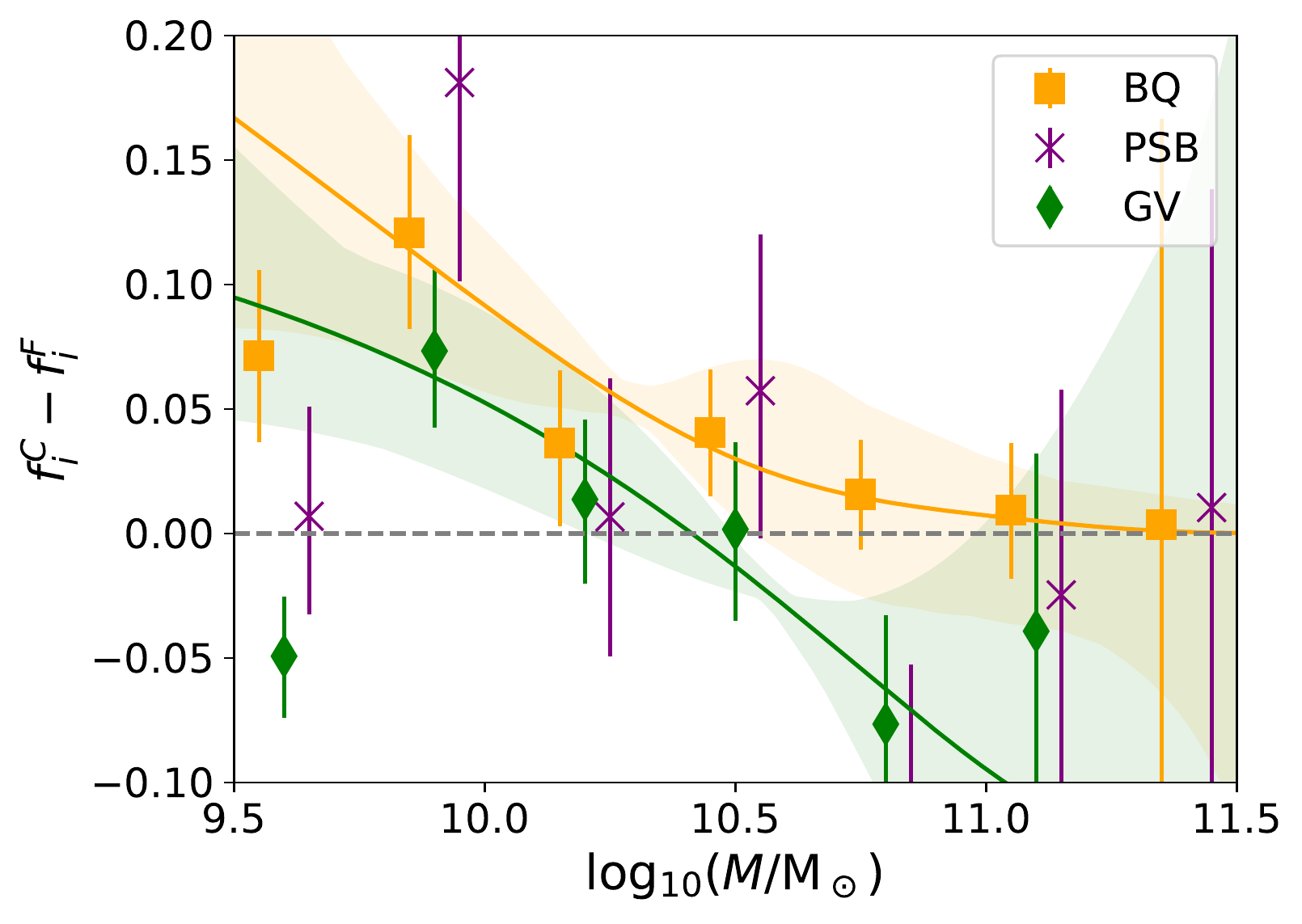} 
    \caption{We show the difference in the abundance of candidate transition galaxies in clusters, relative the field, as a function of stellar mass.   Points show the median with 68 per cent confidence limit error bars, and the corresponding Schechter function fits to the GV and BQ SMFs are shown as solid lines with a shaded region for the 68 per cent confidence limits.  The horizontal, dashed line indicates identical abundances in both environments.  The PSB and BQ points are shifted slightly in the horizontal direction for clarity only.
    }
    \label{fig:GVE_new}
\end{figure}

In Figure~\ref{fig:GVE_new} we show the difference in the fraction of each candidate transition galaxy type in the cluster, relative to the field, as a function of stellar mass.  
For all three populations, this difference is consistent with zero for galaxies with $M>10^{10.5}\,\mathrm{M}_\odot$, but   
increases with decreasing stellar mass, reaching $\sim 10$ per cent at $M\approx 10^{9.5}\,\mathrm{M}_\odot$. 

\subsubsection{Clustercentric radius variations}
\label{section- radial}
\begin{figure}
    \centering
    \includegraphics[scale=0.58]{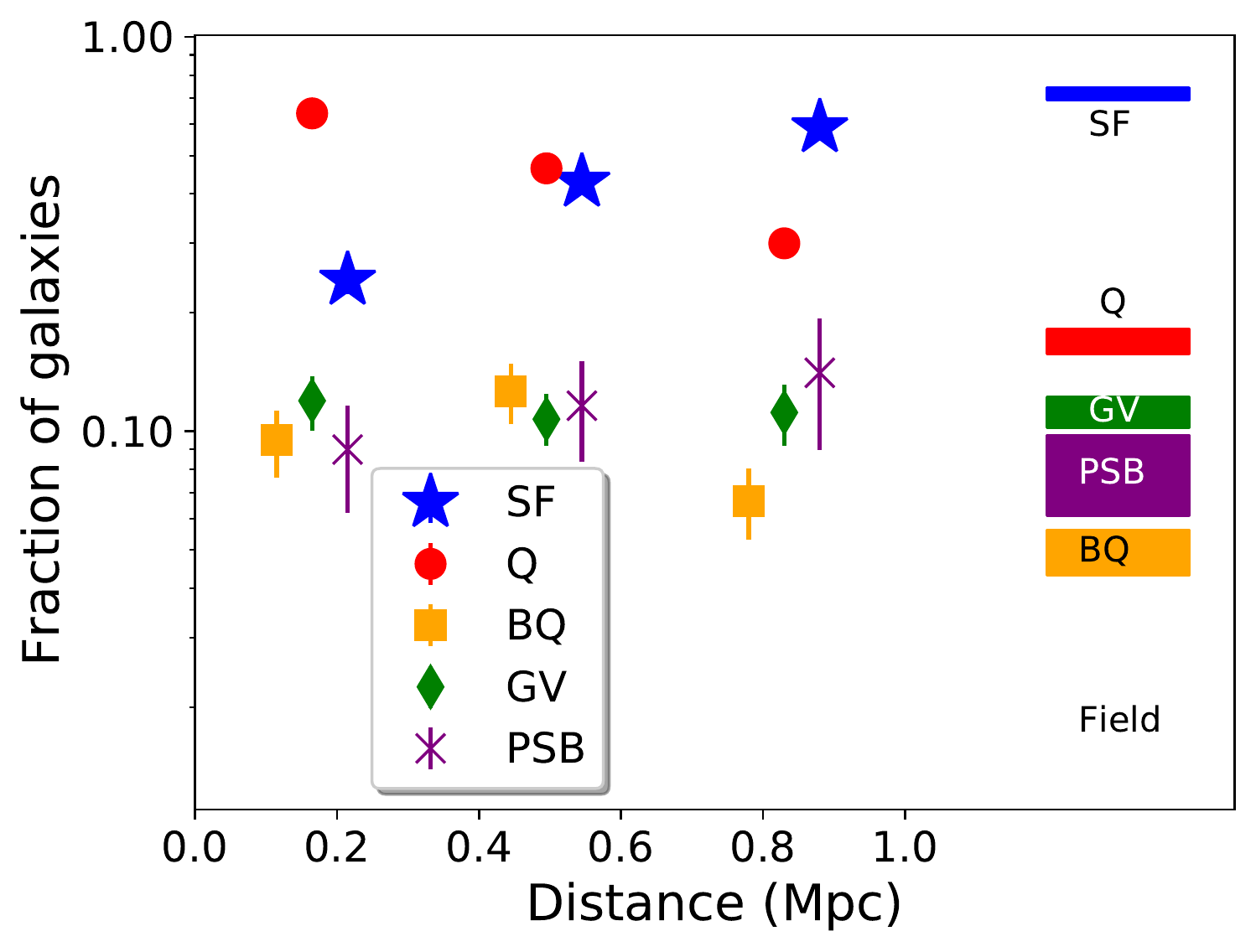} 
    \caption{The relative abundances of each galaxy population considered in this paper are shown as a function of clustercentric distance.  The comparison field is shown as the bars on the right, and includes all the galaxies at $1<z<1.5$ beyond $1.5$ Mpc.  The height of the bars represents the 1$\sigma$ uncertainty on the fractions.
    For the PSBs the plotted fraction is their abundance relative to the total spectroscopic sample, limited to $M>10^{9.5}\,\mathrm{M}_\odot$.  This is well below the spectroscopic completeness limit, and no completeness corrections are applied.  For all the other populations the fraction is relative to the total photometric sample, completeness corrected to $M>10^{9.5} \,\mathrm{M}_\odot$. The PSB and BQ populations are not independent of the other three, so the fractions do not add to unity in a given bin.}
    \label{fig:fracoftotal}
\end{figure}

Figure \ref{fig:fracoftotal} shows how the fraction of galaxies in each population varies with distance from the cluster centre, and how this compares with the corresponding fraction in the field.  These fractions again include all galaxies with $M>10^{9.5}\,\mathrm{M}_\odot$, with completeness weights applied to the photometric samples as described in \S~\ref{sec-sample}.  The PSB population includes galaxies below the spectroscopic completeness limit and we apply no corrections; therefore these values are not necessarily representative of a mass-limited sample.  
As expected, we see an increase in star-forming galaxies with radius, and a decrease in the fraction of quiescent galaxies.  Overall the cluster population shows a highly significant excess of quiescent galaxies, and a corresponding deficit of star-forming galaxies, relative to the field.  

All three transition populations have an abundance of about ten per cent in the cluster environment, with at most modest dependence on distance from the centre.  As noted above, the field fraction of BQ galaxies we measure is consistent with what \citet{belli2019mosfire} find at similar redshifts (at least for the most massive galaxies), and 
significantly lower than what we observe in the cluster population.  However, the fraction of field PSB galaxies is higher than the $\sim 3$ per cent measured by both \citet{muzzin2012} and   \citet{2009MNRAS.398..735Y} at comparable redshifts.  This results in a less significant difference between the cluster and field fractions.
Although the overall fractions of transition galaxies do not show a strong trend with environment, their abundance relative to the SF population does strongly increase toward the cluster centre, as the latter population shows a strong radial trend.    %
Whether or not there is an environmental signature, therefore, depends on what is assumed about the parent population, a point that we will address
in \S~\ref{sec-timescales}.  

\subsubsection{Phase space distribution}\label{phase}
\begin{figure}
    \centering
    \includegraphics[scale=0.55]{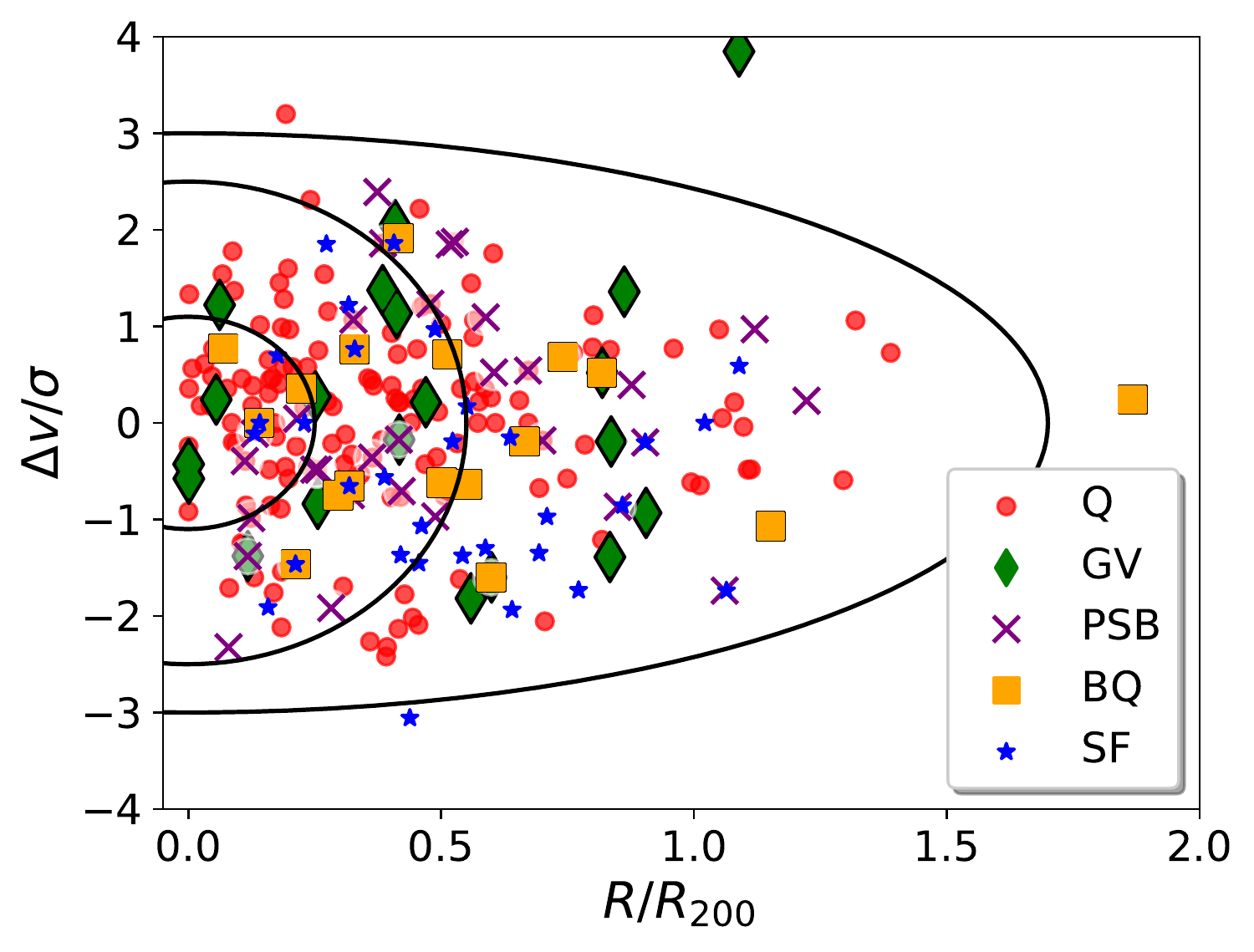}
    \caption{The distribution of galaxies with $M>10^{10.2}\,\mathrm{M}_\odot$ and spectroscopic redshifts in phase space, as defined by  \citet{muzzin2014phase}. The solid contours represent the same phase space bins adopted by \citet{muzzin2014phase}. 
    \label{fig:anya}}
\end{figure}

For the spectroscopic sample we have radial velocity information, so we can also consider their distribution in  position-velocity "phase space", shown in Figure~\ref{fig:anya}.  A similar analysis was done for PSB galaxies in the GCLASS sample of clusters, at slightly lower redshift, by \citet{muzzin2014phase}.  Here we aim to compare directly with that work, so in Figure~\ref{fig:anya} we normalize the clustercentric distances by $R_{200}$, though our conclusions do not change if we use physical distances as in the rest of the paper, due to the modest variation in $R_{200}$ among our cluster sample.  We only consider galaxies more massive than the spectroscopic limit, $M>10^{10.2}\,\mathrm{M}_\odot$, and we overlay the contours from \citet{muzzin2014phase}.  
Figure~\ref{fig:anya} shows that the PSB, BQ and GV galaxies in our sample are distributed throughout phase space.
In any case it is evident from Figure~\ref{fig:SMF_fracs} that abundances vary more strongly with stellar mass than with location in the cluster.  We will therefore focus on those trends for the remainder of this work.

\section{Discussion}\label{sec-discuss}
We have shown that all three transition populations --- GV, BQ and PSB --- are modestly overabundant in clusters, only for low mass galaxies with $M<10^{10.5}\,\mathrm{M}_\odot$. 
It is remarkable in particular that the excess of the GV population is not larger than that of the BQ or PSB galaxies, which  
require relatively rapid quenching on a timescale $< 1$ Gyr. Since the GV will capture all transitions, whether slow or fast, this result implies that most of the cluster-driven transformation may in fact be via the fast-quenching mode described by \citet{belli2019mosfire}, leading to a BQ and PSB appearance.  
In this section we attempt to quantify this further by estimating the timescales associated with each of these populations, and then discuss the implications in the broader context of galaxy evolution.  The quantities we use here to link the transition galaxy populations with a quenching timescale require a number of assumptions and simplifications, and are derived in Appendix~\ref{app-thoughts}.

\subsection{Transition Timescales}\label{sec-timescales}

\begin{figure}
    \centering
    \includegraphics[scale=0.58]{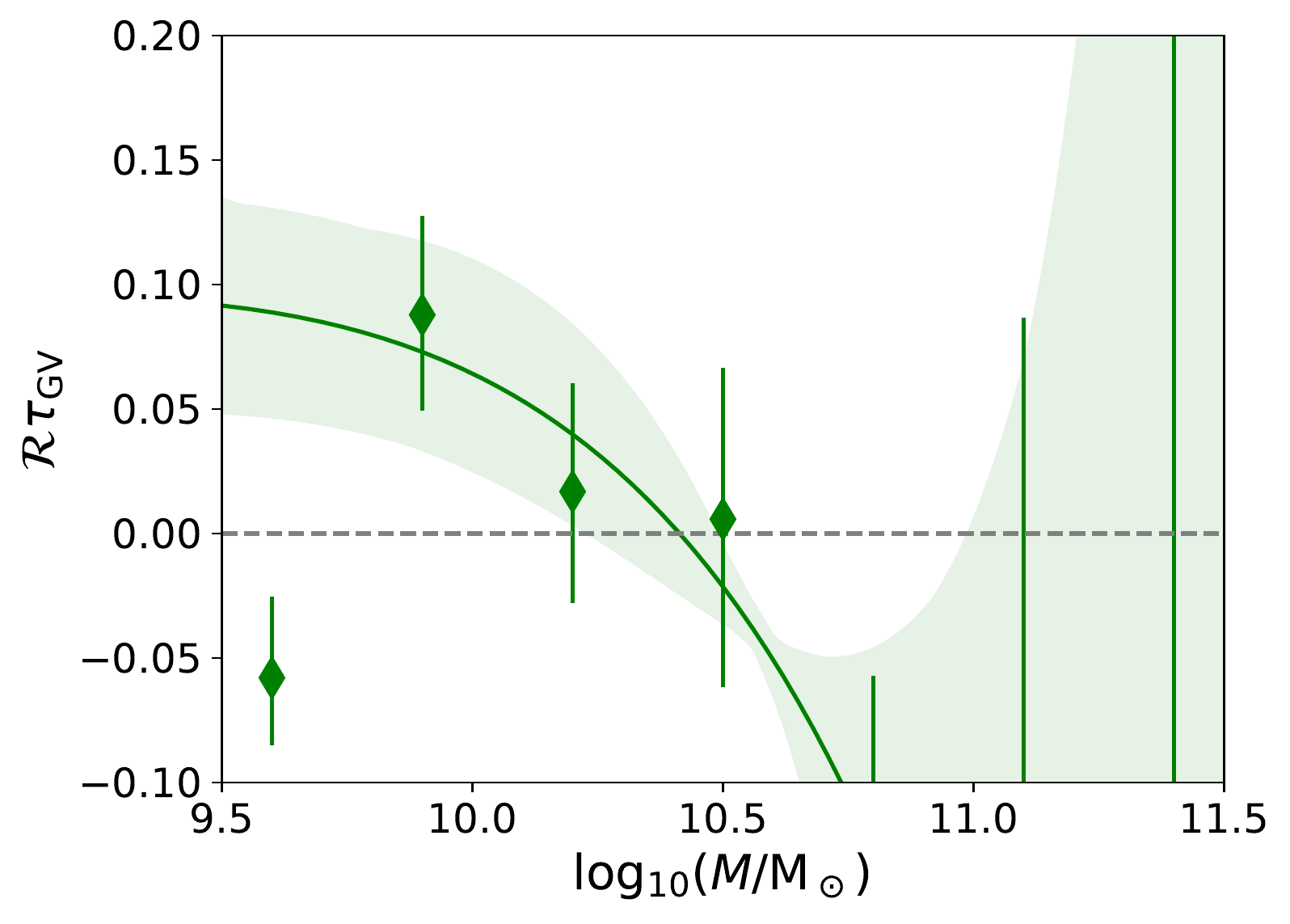} 
    \caption{We show the quantity $\mathcal{R}\tau_{\rm GV}$ as a function of mass, calculated from Equation~\ref{eq-Rtau} for the GV population.  Points show the median with 68 per cent confidence limit error bars, and the corresponding Schechter function fits are shown as solid lines with a shaded region for the 68 per cent confidence limits.  The horizontal, dashed line indicates $\mathcal{R}\tau_{\rm GV}=0$.  Values of $\mathcal{R}\tau_{\rm GV}<0$ correspond to the case where there are fewer cluster GV galaxies than in the field. }
    \label{fig:RTgv}
\end{figure}

The excess abundance of transition galaxies is related to both the rate of environmental quenching, and the time spent in that phase.
As derived in Appendix~\ref{app-trans}, for a given transition population $i$ we can write
\begin{equation}\label{eq-Rtau}
    \mathcal{R}_i\tau_i\approx\frac{f^C_i-f^F_i}{f_{SF}^F},
\end{equation}
where $\mathcal{R}_i$ is the fraction of infalling field SF galaxies that are quenched per unit time, and $\tau_i$ is the time spent in the transition phase $i$.  
This requires a number of assumptions, including that the mass accretion rate of the cluster is constant with time, the excess abundance is due only to quenching (and not, for example, rejuvenation of quiescent galaxies), and that the number of transition galaxies arising due to non-environmental reasons is proportional to the total galaxy population.  Results corresponding to an alternative choice for the latter assumption are presented in Appendix~\ref{app-QFE_R}.
We show $\mathcal{R}_i\tau_i$ as a function of stellar mass for the GV population in Figure~\ref{fig:RTgv}.
As expected given the results shown in Figure\,\ref{fig:GVE_new}, the data are consistent with $\mathcal{R}_{\rm GV}\tau_{\rm GV}=0$ for all stellar masses, apart from one bin.  For the low mass galaxies $M<10^{10.5}\,\mathrm{M}_\odot$ we find a 1$\sigma$ upper limit of $\mathcal{R}_{\rm GV}\tau_{\rm GV}<0.03$. Note that the fit to the data shows a more significant detection, as the lowest mass bin was not included in the fit to the SMFs, as described in \S~\ref{section- mass}.  Values of $\mathcal{R}_i\tau_i<0$ are permitted, and correspond to the case where there are relatively more transition galaxies in the field than the cluster population.  Such values point to limitations in the simple model (for example, neglecting mergers and tidal disruption of galaxies), but we note that we do not measure a statistically significant negative value in any bin.

\begin{figure}
    \centering
    \includegraphics[scale=0.55]{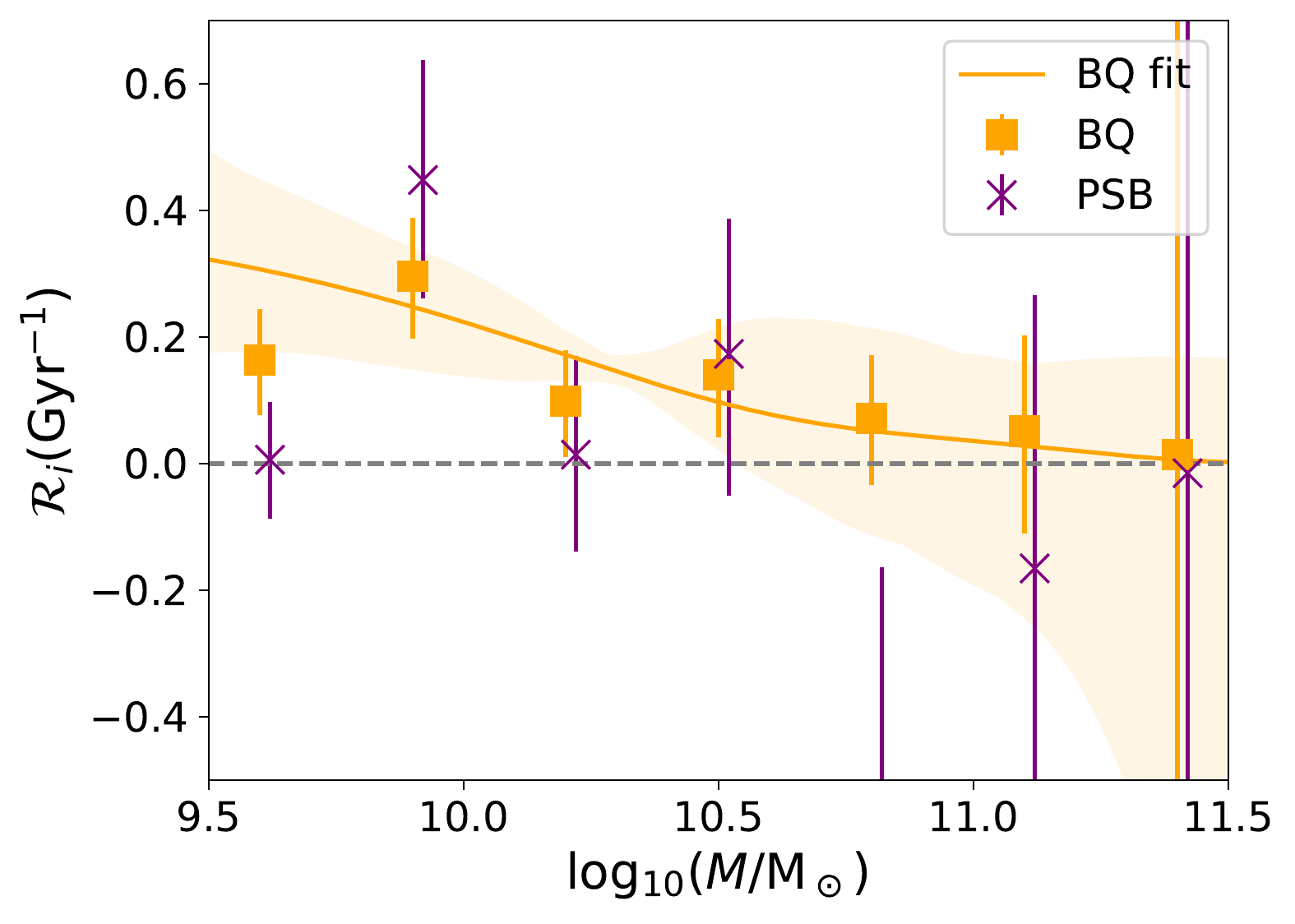}
    \caption{We show the transition rate $\mathcal{R}$ for the BQ and PSB populations, assuming each has a lifetime of $\tau=0.5$ Gyr, as a function of stellar mass.  Points show the median with 68 per cent confidence limit error bars, and the corresponding Schechter function fits are shown as solid lines with a shaded region for the 68 per cent confidence limits.  The PSB points are shifted slightly in the horizontal direction for clarity only.  The horizontal, dashed line indicates $\mathcal{R}_i=0$.  Values of $\mathcal{R}_i<0$ correspond to the case where there are fewer cluster transition galaxies than in the field. }
    \label{fig-Ratebqg}
\end{figure}

\begin{figure}
    \centering
    \includegraphics[scale=0.55]{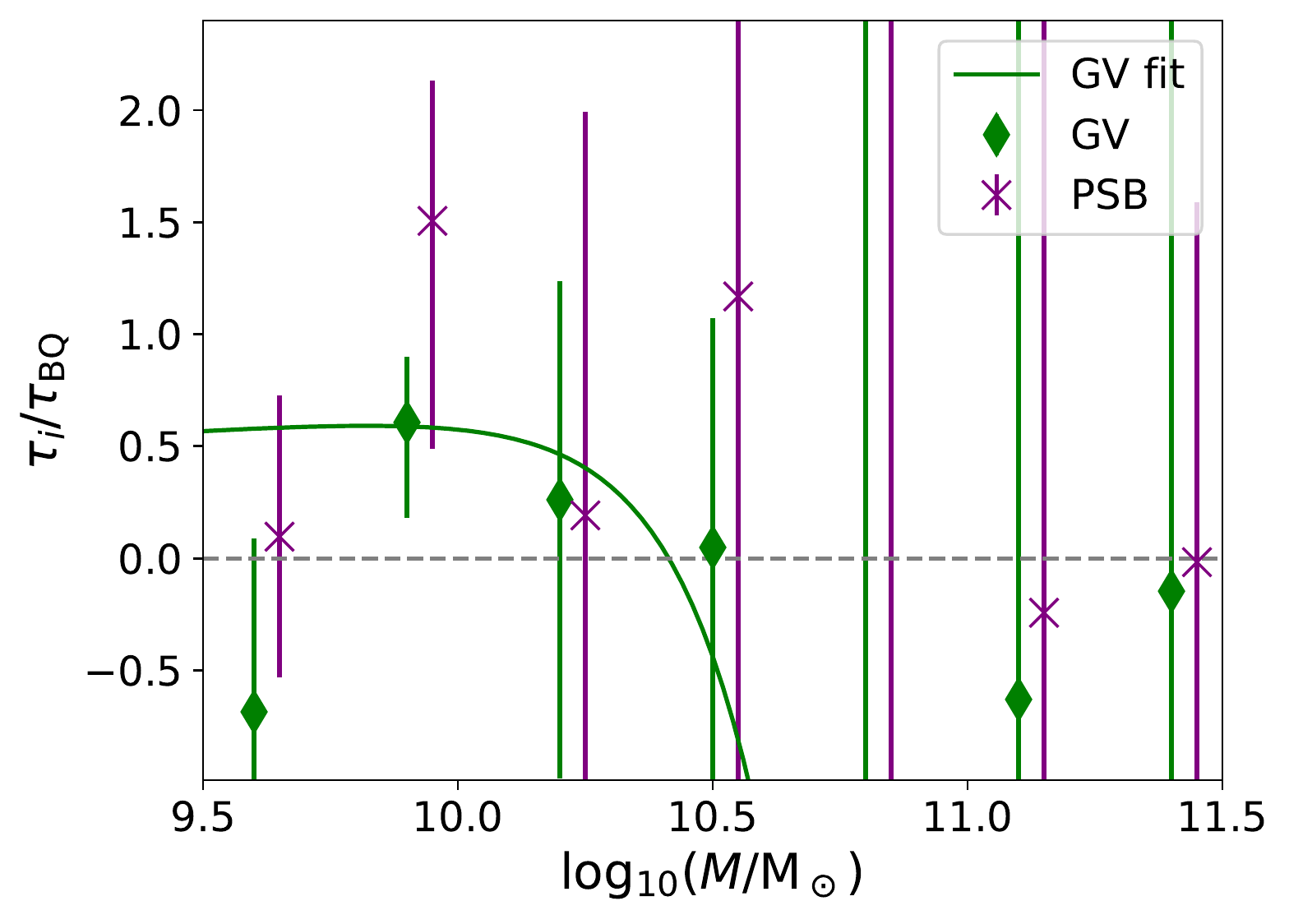}
    \caption{The timescales in each of the PSB and GV phase are shown, relative to the BQ timescale, as a function of stellar mass.  Points show the median with 68 per cent confidence limit error bars, and the corresponding Schechter function fits are shown as solid lines with a shaded region for the 68 per cent confidence limits.  The horizontal, dashed line indicates $\tau_i/\tau_{\rm BQ}=0$.   The PSB points are shifted slightly in the horizontal direction for clarity only.  }
    \label{fig-tauratios}
\end{figure}

 From analysis of the spectra of both BQ \citep{belli2019mosfire} and PSB \citep{muzzin2012,French18} galaxies it has been argued that their lifetimes are $\tau\sim 0.5\pm 0.3$ Gyr.  If we adopt this timescale (neglecting the large uncertainty for the moment) we can estimate the effective quenching rate $\mathcal{R}$ over that time from Equation~\ref{eq-Rtau}.
These rates are shown, as a function of stellar mass, in Figure~\ref{fig-Ratebqg}.
The uncertainties are too large to permit strong conclusions, even before considering the additional uncertainty in assumed timescale $\tau$.  But both populations yield similar results and are consistent with a quenching rate at the epoch of observation that is significantly greater than zero, and consistent with each other to within a factor of two, for low mass galaxies.
The median value for the low mass BQ sample, $M<10^{10.5}\,\mathrm{M}_\odot$, is $\mathcal{R}_{\rm BQ}=0.17\pm0.05\,\mbox{Gyr}^{-1}$.   For the PSB galaxies we find $\mathcal{R}_{\rm PSB}=0.09\pm0.07\,\mbox{Gyr}^{-1}$, though we remind the reader that the spectroscopic sample on which this is based is not representative of all galaxy types for $M<10^{10.2}\,\mathrm{M}_\odot$.  Adopting a longer (shorter) timescale $\tau_{i}>0.5$ would reduce (increase) these rates proportionally. 

As the data only constrain the combination $\mathcal{R}_i\tau_i$, we cannot determine timescales $\tau_i$ directly without an assumption about $\mathcal{R}_i$.  Under the assumption that the three populations are different phases of the same quenching process, then their corresponding rates should be equal,
$\mathcal{R}_{BQ}=\mathcal{R}_{PSB}=\mathcal{R}_{GV}$.
In this case we can calculate their relative timescales.  We show the GV and PSB timescales relative to that of the BQ population, in Figure~\ref{fig-tauratios}.  We choose the BQ as the reference population because it yields the best constraints on $\mathcal{R\tau}$, due to its relatively large sample size (compared with the spectroscopic PSB sample) and its contrast with the field (compared with the GV galaxies). From this we conclude that, for low mass galaxies, the lifetimes in the PSB and GV phase are consistent with the lifetime of the BQ phase.  If anything, the GV abundance requires an even shorter lifetime, $\tau_{\rm GV}=(0.3\pm 0.3)\tau_{\rm BQ}$, suggesting that slow-quenching through the green valley is not a dominant mode for environmental transformation. For example, the slow quenching model of \citet{belli2019mosfire} predicts $UVJ$ colours that are consistent with our GV definition for $\sim 1$ Gyr (comparing their Figure 12 with the right panel of our Figure~\ref{fig:photselect}).  This does not, however, rule out the possibility of a long delay time between when a galaxy first enters a dense environment and when quenching begins \citep[e.g.][]{wetzel2012galaxy,DeLucia2012,mok2014formation,MBB,Foltz}.  
From an analysis of the SF population in GOGREEN, \citet{old2020gogreen} measured a small but significant difference in the SFRs of cluster galaxies relative to the field, in the sense that low mass cluster SFRs are lower by about 50 per cent.  As these galaxies are independent of the transition galaxies studied here, this implies that there is a modest environmental effect present prior to the rapid transition through the GV/BQ/PSB phase, as observed at lower redshift \citep[e.g.][]{Vulcani10,Haines15,Finn2018}.

\subsection{The evolution of cluster galaxy quenching}\label{sec-discuss_GG}

It is notable that the QFE we \citep[and][]{van2020gogreen} measure, $\sim 0.4$, is very similar to what is measured at $z\sim 0$ \citep{peng2012,2014MNRAS.440..843O} and at $0.4<z<0.8$ \citep{poggianti2009environments} for similar environments.   Given the shorter age of the Universe at $z=1$, we show in Appendix~\ref{app-thoughts} that a time independent QFE implies (under several assumptions) that the quenching rate must decrease with time as $\mathcal{R}\propto t^{-1}$, and consequently $\mathcal{R}=T\times QFE$, where $T$ is the lifetime of the cluster at a given epoch.
\begin{figure}
    \centering
    \includegraphics[scale=0.52]{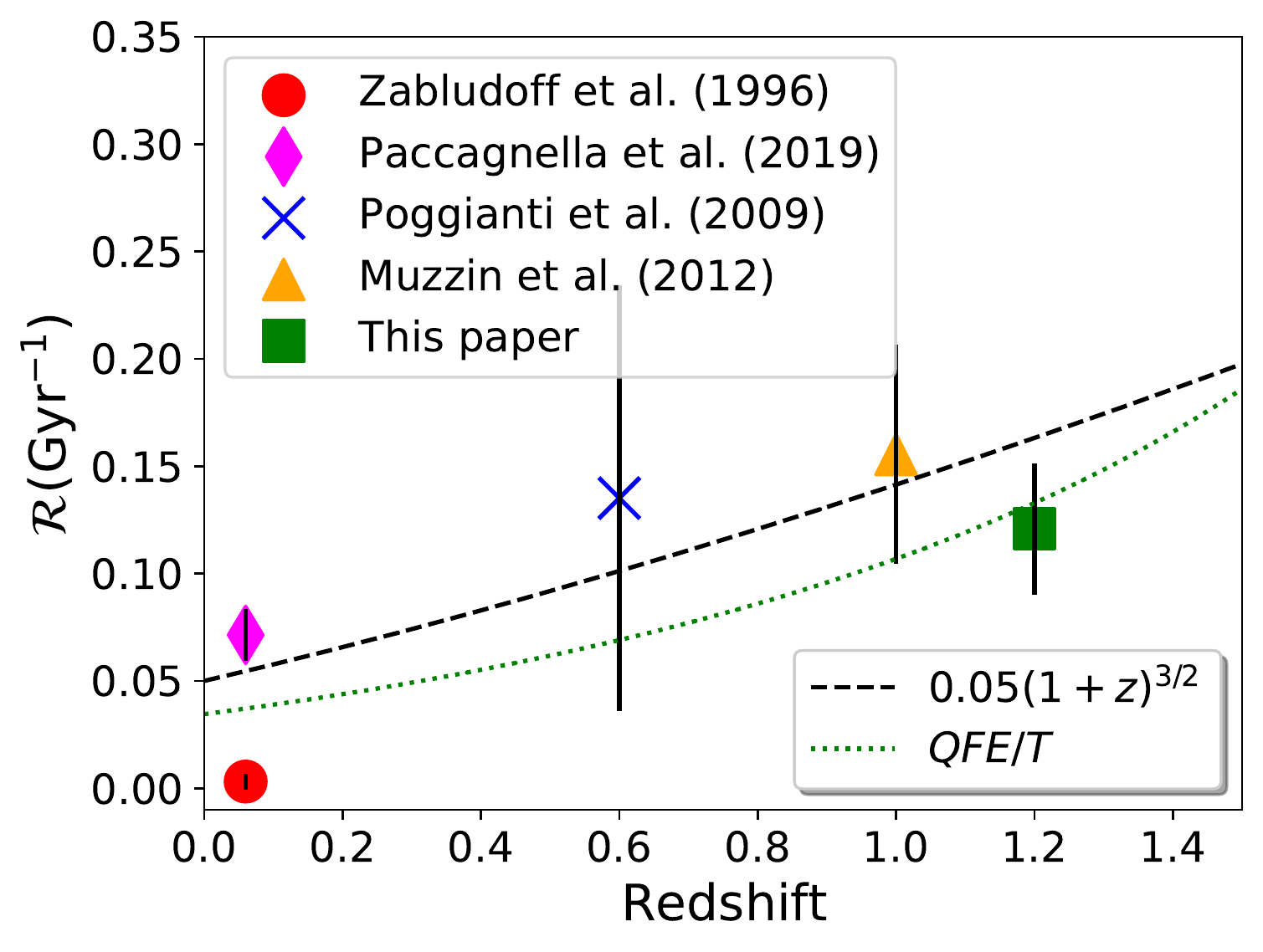}
    \caption{The rate of cluster-driven quenching, is calculated from Equation~\ref{eq-Rtau} for the BQ population with $M\gtrsim 10^{10}\,\mathrm{M}_\odot$
    at three redshifts, from the sources indicated in the legend.   We assume that $\tau_{\rm PSB}=0.5$ Gyr is constant.  The dashed line shows an arbitrarily normalized timescale that evolves inversely with the dynamical time.  The dotted line represents the rate evolution implied by a constant QFE=0.4 (Appendix~\ref{app-thoughts}), assuming the cluster forms at $z=3$.  
    }
    \label{fig-PSBevol}
\end{figure}

We can therefore use observations of post-starburst galaxies in clusters at different redshifts to estimate the evolution of $\mathcal{R}$ directly from Equation~\ref{eq-Rtau}, and compare this with the predicted evolution assuming a constant QFE. 
This is shown in  Figure~\ref{fig-PSBevol}, where we compare our results with the abundance of spectroscopic PSB in massive clusters observed at $z\sim 0$ from \citep{zabludoff1996environment} and \citet{paccagnella2019strong}, and at intermediate redshifts from \citet{poggianti2009environments}, assuming $\tau=0.5$ Gyr is constant.   
The literature values are based on the fraction of PSB galaxies relative to the total population in luminosity-limited samples that are comparable to the stellar mass limit of the present work, though not precisely matched\footnote{Given the strong mass-dependence we find, a more careful analysis in future work is warranted.}.  
For GOGREEN, the data point shown is based on the BQ population, as this is complete to lower masses than the spectroscopic PSB sample, but implies rates that are consistent with those obtained from the PSB, with smaller uncertainties.
The data show a modest evolution of $\mathcal{R}$ over $0<z<1.2$, strongest when compared with the low fraction of PSB observed by \citet{zabludoff1996environment} at $z\sim 0$ \citep[see also][]{hogg2006triggers,Vulcani15}.  This is compared with the expected evolution implied by a constant QFE, $\mathcal{R}=T\times QFE$ where $T$ is computed 
assuming a formation redshift $z_f=3$.  Given the substantial statistical and systematic uncertainties, the data are consistent with this simple model, and hence with the observation that QFE is constant.   We also compare with the evolution of the inverse of the dynamical time, which has similar behaviour that is indistinguishable given these data.

\begin{figure}
    \centering
    \includegraphics[scale=0.52]{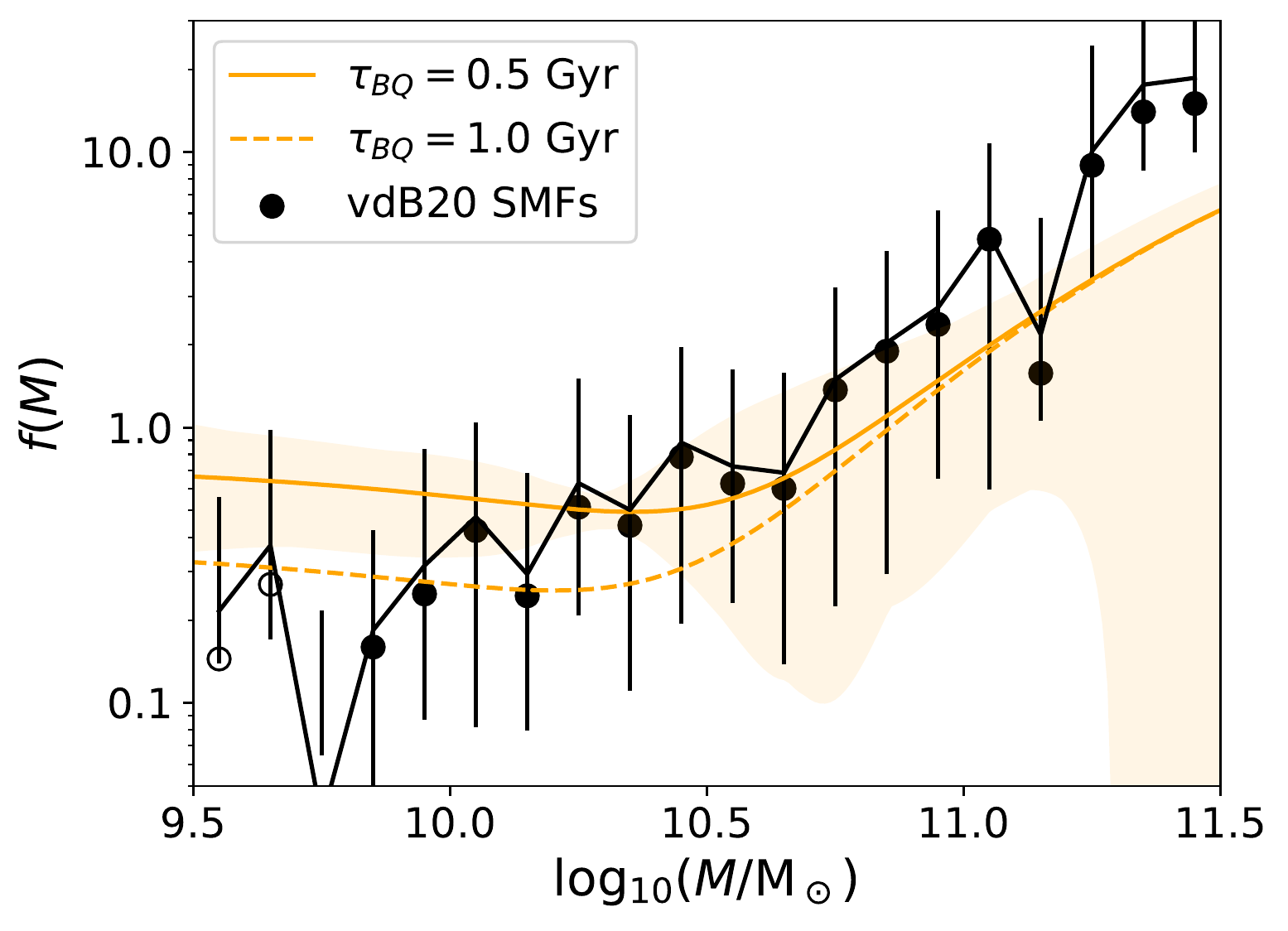}
    \caption{We show the fraction of field SF galaxies that have to be added to the quiescent population to match the observed cluster quiescent SMF, as described in Appendix~\ref{app-QFE_fM}.  The points with error bars use the SMFs of \citet{van2020gogreen}; the open symbols at the lowest stellar mass require large completeness corrections, which have been applied.  Note that $f(M)\neq QFE$ because the total SMF shape is environment dependent.  The orange curves show the $f(M)$ using the QFE that is inferred from the excess BQ population,  assuming two different timescales $\tau_{\rm BQ}=1$ and $0.5$ Gyr as indicated.   We show the 68 per cent confidence limits, for the $0.5$ Gyr case only, as the shaded region.
    }
    \label{fig-fM}
\end{figure}

Analysing the same GOGREEN clusters, with the same data, \citet{van2020gogreen} showed that the shape of the quiescent galaxy SMF is identical in the cluster and field environments.  This is significantly different from what is found at low redshift \citep[e.g.][]{baldry2006galaxy,peng2012}, where the low-mass end of the quiescent SMF in clusters rises steeply with decreasing mass, as would be expected if SF galaxies accreted from the field were being quenched independently of their stellar mass. In particular, \citet{van2020gogreen} showed that the shape of the cluster quiescent SMF cannot be reproduced simply by adding a mass-independent fraction $f$ of field SF galaxies to the quiescent population, as
\begin{equation}\label{eq-def_f}
    \phi_Q^C(M)=A\left[\phi_Q^F(M)+f\phi_S^F(M)\right],
\end{equation}
where $A$ is a renormalization constant.  In Appendix~\ref{app-QFE_fM} we show that Equation~\ref{eq-def_f} holds if $f$ is a function of mass.  Specifically, $f(M)$ can be related to the QFE and to the relative shapes of the total SMFs, as:
\begin{equation}
    f(M)=\xi(M)QFE(M)+\frac{\phi_Q^F(M)}{\phi_S^F(M)}\left[\xi(M)-1\right],\tag{\ref{eq-f_QFE}}
\end{equation}
where $\xi(M)=\phi^C(M)/A\phi^F(M)$.
Note that $f\neq QFE$ unless the total SMF in the cluster and field have the same shape, so that $\phi^C=A\phi^F$.  Instead we obtain two terms.  The first is due to converting SF galaxies to Q, and characterised by the QFE.  The second term results from needing to add an excess of both SF and Q galaxies from the field to account for the different total SMF shape.  We show this function $f(M)$ in Figure~\ref{fig-fM}, using the SMFs from \citet{van2020gogreen}.  
 We find that $f(M)$ has to be strongly mass dependent, exceeding unity for $M>10^{10.75}\,\mathrm{M}_\odot$.  The latter point indicates that the model in Equation~\ref{eq-def_f} is not appropriate for the highest mass galaxies, as they cannot be obtained simply by quenching SF field galaxies - there are not enough of them.  There must be an additional source, for example through merging of lower mass galaxies, or preferential accretion of massive galaxies.  

We can compare this constraint on $f(M)$ with independent predictions of the QFE, via the quenching rate $\mathcal{R}$ derived from the BQ galaxy abundance.  
This depends on the BQ lifetime, and we will show the results assuming both $\tau_{BQ}=0.5$ Gyr and $\tau_{BQ}=1.0$.  From this we can predict the first term in Equation~\ref{eq-f_QFE} for $f(M)$, again assuming $\mathcal{R}=T\times QFE$, and a formation redshift $z_f=3$.  The result is shown as the orange curves in Figure~\ref{fig-fM}.  At the high mass end the shape is dominated by the second term in Equation~\ref{eq-f_QFE}, that is independent of quenching and hence the BQ constraints.  The predicted curve has a similar shape to the data, but lies below it.  This difference reflects our observation that the QFE inferred from the SMFs is large at these masses, yet we measure no significant excess of transition galaxies.  Within the uncertainties, however, the two estimates of $f(M)$ are fully consistent.
At low stellar masses, the flat trend and high normalization (for $\tau_{\rm BQ}=0.5$ Gyr ) inferred from $\mathcal{R}_{\rm BQ}$ contrasts with that obtained directly from the SMFs.  Adopting the longer timescale of 1\,Gyr is sufficient to bring them into agreement; but such a long lifetime is at the upper end of the expectations from their ages \citep{belli2019mosfire}, and also would imply quenching rates $\mathcal{R}$ a factor of two below what we show in Figure~\ref{fig-PSBevol}, and therefore in contrast with expectations from a non-evolving QFE.  There are substantial statistical and systematic uncertainties, as well as numerous assumptions described in Appendix~\ref{app-thoughts}.  However, it appears that a more complex model may be required to reconcile the large QFE at low stellar masses, which is supported by the observed abundance of transition galaxies, with the shallow slope of the quiescent SMF in clusters that drives the low $f(M)$ in Figure~\ref{fig-fM}. 
The fact that the SMF of the BQ population (and, indirectly, the PSB population) looks similar to that of SF galaxies in the field further reinforces that we would expect to see an excess of low mass galaxies in the quiescent SMF, that does not appear to be there.  

In another work studying these same clusters,  \citet{Webbgogreen} found that the mass-weighted ages of cluster and field quiescent galaxies are comparable; if anything, cluster galaxies are somewhat older.  This is not what would be expected if cluster galaxies were first quenched upon infall, as these environments would then host quiescent galaxies that are younger than those in the field \citep[e.g.][]{muzzin2012}.  A proposed interpretation is that "pre-processing" is important; i.e. environmental quenching is initiated when galaxies first become satellites, long before they are accreted into more massive clusters.  Analysis of GOGREEN systems as a function of halo mass supports this interpretation for massive galaxies \citep{Reevesgogreen}, as does a recent analysis of galaxies in the infall regions of these same clusters (Werner et al. submitted).  These observations, together with the older ages measured by \citet{Webbgogreen} and our results here showing no significant excess of transition galaxies for massive galaxies, all point to the need for these galaxies to have ended their star formation at $z>2$, possibly in a protocluster environment, and that accretion onto the existing cluster at $z\sim 1$ plays little or no role in further star formation quenching.

The picture is a little less clear for the lower mass galaxies.  Here, the abundance of transition galaxies is enough to fully explain the QFE under simple assumptions, suggesting that accretion-driven quenching is the dominant route for forming quiescent galaxies.  This is consistent with the work of \citep{Reevesgogreen} and Werner et al. (submitted), which indicate little or no pre-processing in this mass regime.  It is in some tension  with the shape of the quiescent SMF (Figure~\ref{fig-fM}), and the small age difference of $0.1^{+0.46}_{-0.37}$ Gyr between cluster and field quiescent galaxies found by  \citet{Webbgogreen}.  However, the statistical and systematic uncertainties in both cases are large enough that we cannot rule out a simple interpretation where accretion-driven quenching plays an important or even dominant role for creating low mass, quiescent galaxies in clusters at $z\sim 1$. 
Further insight will be gained from a careful study of how the SMFs of these different populations depend on halo mass and redshift \citep[e.g.][]{wild2020formation,socolovsky2018enhancement}.  Moreover, observations of protoclusters at $z\gtrsim 2$ will help to determine the "initial conditions" of the cluster population, to disentangle accretion-driven quenching from differences that arise in the early cluster formation phase.

\section{Conclusions}\label{sec-conc}
We have taken advantage of deep rest-frame $NUV$ photometry in the GOGREEN sample of $1<z<1.5$ galaxy clusters to identify transition galaxies in the green valley (GV) between the dominant star-forming and quiescent populations.  This definition has the advantage of being complete, in the sense that it is certain to include every galaxy that transitions from the star-forming to quiescent population, at some point during its evolution.  
We compare this population with two other definitions of transition galaxies - blue quiescent (BQ) galaxies and spectroscopically identified post-starburst (PSB) galaxies.  These have the advantage that they are likely more pure, in the sense that a larger fraction of the sample are expected to be in the process of {\it rapidly} quenching their star formation.  They may be incomplete, as it is possible for galaxies to transition without passing through either of these phases.  
Our main observations are summarized as follows: 

\begin{itemize}
    \item The SMF of GV galaxies has a shape most similar to that of the Q population, and inconsistent with that of the star-forming (SF) galaxies. The opposite is true of the BQ galaxies.  Cluster BQ galaxies have strong H$\delta$ absorption in their spectra, $5\pm 1$ \AA\ equivalent widths, comparable to that of the PSB galaxies.   The three transition populations partially overlap in colours and D4000 break strength, but are substantially independent.  They may sample different phases of evolution during the final quenching of star formation, or different pathways taken by different galaxies depending on their dust evolution and past star formation history.
    \item For massive galaxies, $M>10^{10.5}{\rm M}/M_\odot$, we find no significant difference in the fraction of PSB, BQ or GV galaxies in clusters, compared with the field.
    \item For low stellar masses $M<10^{10.5}{\rm M}/M_\odot$, BQ, PSB and GV galaxies make up a $\sim 5-10$ per cent greater proportion of the cluster population than in the field. 
    \item We do not find a strong radial or phase space dependence on the distribution of transition galaxies relative to the total population.  Relative to the SF population, all three populations are significantly more abundant in the central regions of the cluster.  
    \item Assuming the PSB and BQ phases last for $\sim 0.5$ Gyr, as argued by others based on their spectra \citep{belli2019mosfire,muzzin2014phase,French18}, their abundances imply that 20--30 per cent of infalling SF galaxies with $M<10^{10.5}\,\mathrm{M}_\odot$ are quenched every Gyr.
    \item The relative abundance of GV galaxies implies environmentally quenched galaxies exist in this phase for a comparable or even shorter time than in the BQ phase.  
\end{itemize}

We conclude that  the BQ population is the best tracer of recent environmental quenching, as it i) can be identified photometrically; ii) exhibits the largest excess abundance in clusters, relative to the field; iii) has a SMF consistent with the SF population, and average spectra that are most similar to classic post-starburst spectra, with strong H$\delta$ absorption; and iv) appears to capture most, if not all, of the environmentally quenching galaxies at some point in their lifetime.  The GV, on the other hand, shows at most a modest excess abundance in clusters relative to the field, and appears to be the most heterogeneous, with environmentally quenching galaxies comprising less than half the cluster population.  

Our results are consistent with a fairly simple picture that is qualitatively different for high and low mass galaxies.  For massive galaxies, $M>10^{10.5}\,\mathrm{M}_\odot$, we do not find a significant excess of any type of transition galaxy -- GV, BQ or PSB -- in clusters.  Though the uncertainties are too large to draw strong conclusions, the results are consistent with most massive cluster quiescent galaxies being quenched before accretion on to the cluster, via pre-processing in group or protocluster environments \citep[][Werner et al., submitted]{Reevesgogreen}. 
For lower mass galaxies, $M<10^{10.5}\,\mathrm{M}_\odot$,
the excess of GV, PSB and BQ galaxies all indicate that approximately 20--30 per cent of the star-forming cluster population is quenched every 1~Gyr, in excess of what would be expected in the field.  The final quenching process itself is rapid ($\tau<1$~Gyr), and most quenching galaxies pass rapidly through the GV and go through both a PSB phase (with blue colours but no nebular emission) and a BQ phase (at the blue end of the quiescent region of the $UVJ$ diagram).  The inferred quenching rate is consistent with the observed QFE at both $z=1$ and $z=0$, implying that this rapid quenching can account for most of the low mass excess quiescent population in clusters. This does not exclude a longer delay time preceding the final quenching, that can lead to a modest reduction in SFR \citep[e.g.][]{old2020gogreen} prior to the rapid transition.  
We conclude therefore that most $M<10^{10.5}\,\mathrm{M}_\odot$ quiescent galaxies were quenched upon infall into the cluster.

\newpage
\section*{Data Availability}
All data used in this paper are available from the GOGREEN and GCLASS public data release, at 
the CADC (\url{https://www.cadc-ccda.hia-iha. nrc-cnrc.gc.ca/en/community/gogreen}), and NSF’s NOIR-Lab (\url{https://datalab.noao.edu/gogreendr1/}). 

\section*{Acknowledgements}
We thank the native Hawaiians for the use of Maunakea, as observations from Gemini, CFHT, and Subaru were all used as part of our survey.

Data products were used from observations made with ESO Telescopes at the La Silla Paranal Observatory under ESO programme ID 179.A-2005 and on data products produced by TERAPIX and the Cambridge Astronomy Survey Unit on behalf of the UltraVISTA consortium. As well, this study makes use of observations taken by the 3D-HST Treasury Program (GO 12177 and 12328) with the NASA/ESA HST, which is operated by the Association of Universities for Research in Astronomy, Inc., under NASA contract NAS5-26555.
MB gratefully acknowledges support from the NSERC Discovery Grant program.
BV acknowledges financial contribution  from the grant PRIN MIUR 2017 n.20173ML3WW\_001 (PI Cimatti) and from the INAF main-stream funding programme (PI Vulcani). GW acknowledges support from the National Science Foundation through grant AST-1517863, HST program number GO-15294, and grant number 80NSSC17K0019 issued through the NASA Astrophysics Data Analysis Program (ADAP). Support for program number GO-15294 was provided by NASA through a grant from the Space Telescope Science Institute, which is operated by the Association of Universities for Research in Astronomy, Incorporated, under NASA contract NAS5-26555. GR thanks the International Space Science Institute (ISSI) for providing financial support and a meeting facility that inspired insightful discussions for team “COSWEB: The Cosmic Web and Galaxy Evolution”. GR acknowledges support from the National Science Foundation grants AST-1517815, AST-1716690, and AST-1814159,  NASA HST grant AR-14310, and NASA ADAP grant 80NSSC19K0592. GR also acknowledges the support of an ESO visiting science fellowship. This work was supported in part by NSF grants AST-1815475 and AST-1518257. R.D. gratefully acknowledges support from the Chilean Centro de Excelencia en Astrof\'isica y Tecnolog\'ias Afines (CATA) BASAL grant AFB-170002. J.N. received support from Universidad Andr\'es Bello internal grant DI-12-19/R. KW acknowledges support from NSERC through a CGS-D award.

\bibliographystyle{mnras}
\typeout{}
\bibliography{references}

\appendix

\section{Quenched Fraction Excess and Transition Rates}\label{app-thoughts}
Here we consider how the Quenched Fraction Excess (QFE, Equation~\ref{eq-QFE}) can be related to the instantaneous quenching rate of infalling galaxies.  In particular, we are looking for some insight into the observation that QFE at fixed stellar mass and environment has been approximately constant, over the past 9 Gyr. We will demonstrate that QFE can be linked to a quenching rate as we have assumed in this paper, but that it requires several strong assumptions.  Furthermore, the interpretation is not unique, and we briefly demonstrate two other models for producing an excess of quenched galaxies that can lead to a constant QFE.

As in the main body of the paper, we use $\phi_i^j$ to represent the stellar mass function of population $i\in\{SF,Q,PSB,GV,...\}$ in environment $j\in\{C,F\}$ for cluster or field.  The total SMF in each environment is written without a subscript, $\phi^C$ and $\phi^F$.  Finally we write the fraction of galaxies in population $i$ and environment $j$ as $f_i^j=\phi_i^j/\phi^j$.  

\subsection{Relating QFE to a quenching rate}\label{app-QFE_R}
\begin{figure}
    \centering
    \includegraphics[scale=0.5]{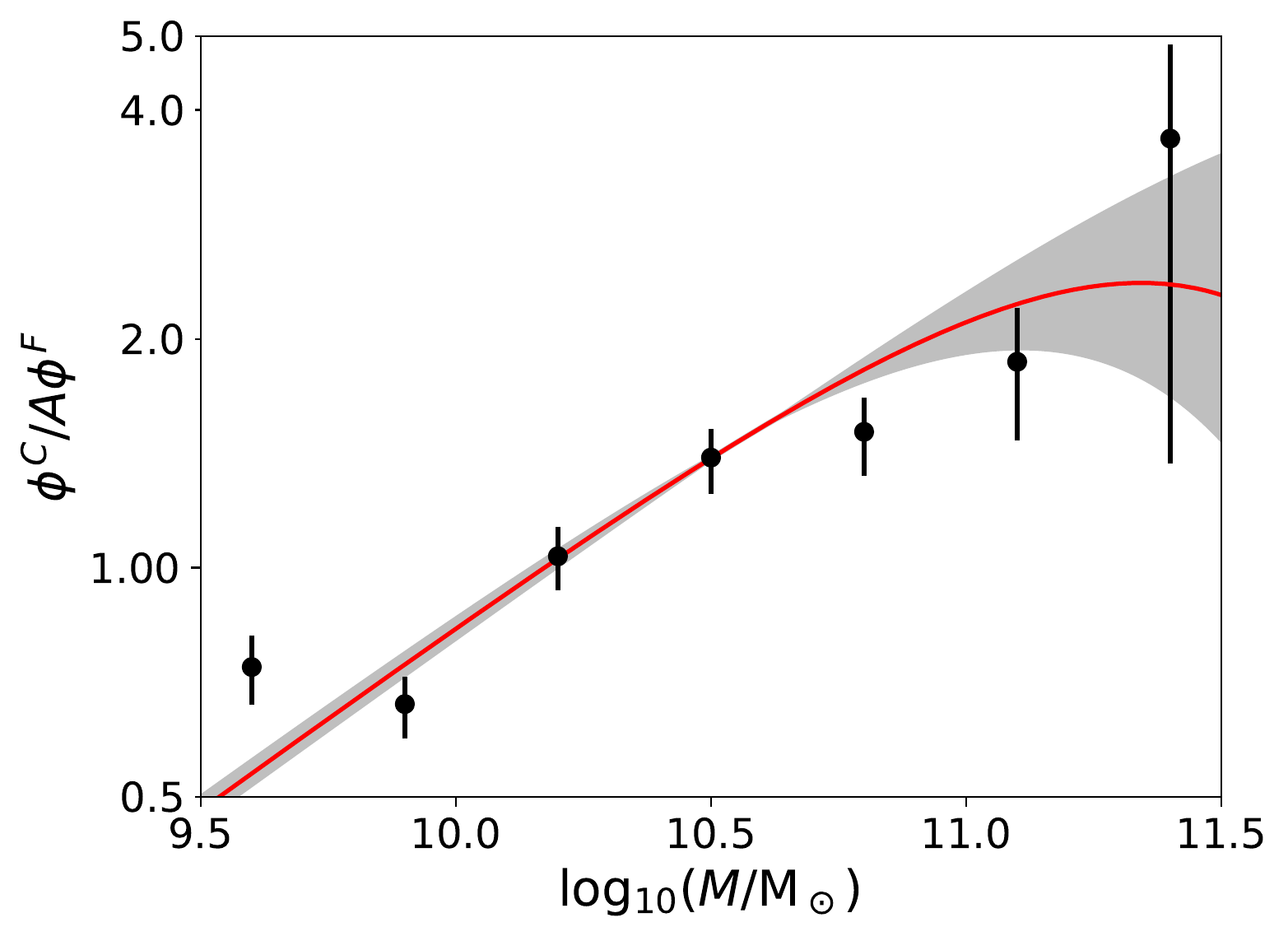} 
    \caption{The ratio of the total stellar mass functions in the cluster ($\phi^C$) and field ($\phi^F$) samples are shown for the binned data (medians with 68 per cent confidence limit error bars) and the Schechter function fits (red line for the best-fit parameters, and grey shaded region for the 68 per cent confidence region).  The field SMF has been renormalized so the integral matches that of the cluster.   The mass dependence reflects the different shapes of the total SMFs, in particular the relative overabundance of massive galaxies in the cluster sample.}
    \label{fig:gamma_new}
\end{figure}

Let $\gamma\phi^F$ represent the number of field galaxies accreted onto the cluster per Gyr.  If $\gamma$ is approximately constant over the lifetime $T$ of a cluster, we can write
\begin{equation}
    \phi^C(M,T)\approx\gamma(M)\,\phi^F(M,T) \,T, 
\end{equation}
or 
\begin{equation}\label{eq-gamma}
    \gamma(M)=\frac{1}{T}\frac{\phi^C(M,T)}{\phi^F(M,T)}. 
\end{equation}
 Thus, $\gamma\propto \phi^C/\phi^F$, and we show this ratio from our observed total SMFs in Figure~\ref{fig:gamma_new}, with the field SMF renormalized by a factor 
 $A=\int{\phi^C(M)dM}/\int{\phi^F(M)dM}$.
 The dependence on stellar mass reflects the fact that the shape of the SMF is environment-dependent, with relatively more massive galaxies in the cluster sample, as also found for this same cluster sample by \citet{van2020gogreen}.  In what follows we will assume that this is due to cluster environments "sampling" a different stellar mass distribution from the average field, but neglect important environmentally-driven mass transformation effects like mergers or tidal disruption.    

Now we assume that the rate $\mathcal{R}$ represents the fraction of infalling SF galaxies that are quenched per Gyr.  Importantly in what follows, we assume $\phi_{\rm SF}^F(T)=\phi_{\rm SF}^F$ is constant, as motivated empirically \citep{ultravista,2021MNRAS.503.4413M}.  $\phi_{\rm SF}^F(T)$ represents galaxies that are still SF in the field at the time of observation, $T$, and the only ones we care about when computing environmental quenching - others have been "self-quenched", for reasons independent of environment. 
Making the strong assumption\footnote{This assumption may be appropriate if accretion is defined as infall onto the main progenitor, but less so if it is the first time a a galaxy becomes a satellite \citet{DeLucia2012}.} that galaxies of all masses are accreted at the same rate, the total number of galaxies accreted between time $0$ and $t$ that are still SF at the time of observation $T$ is
\begin{equation}\label{eq-Sa}
    \phi_{S,a}(M,t)=\int_{0}^{t}{\gamma(t^\prime)\phi_{\rm SF}^F(M,T)dt^\prime}\approx\gamma\phi_{\rm SF}^F(M)t.
\end{equation}
The $\phi_{\rm SF}^F(M,T)$ in the integrand is always evaluated at the same mass $M$ corresponding to $\phi_{S,a}(M)$ on the left hand side: $\gamma\phi_{\rm SF}^F(M)dt$ is therefore the number of galaxies at {\it final} mass $M$ that were accreted in that time interval, whatever their actual mass at earlier time $t$.  Here and below we are making a strong (and almost certainly incorrect) assumption that the quenching rate $\mathcal{R}$ depends only on the final mass $M$ of a galaxy.  
In what follows we drop the explicit $M$ dependence to make the formulas more concise.

The total number of quenched cluster galaxies is given by the sum of the environmentally quenched and self-quenched populations:
\begin{equation}\label{eq-phiQC}
    \phi_Q^C=\phi^C_{Q,env}+\phi^C_{Q,s}.
\end{equation}
Expanding the first term on the right hand side we get
\begin{align}\label{eq-phi_Sa}
    \phi^C_Q=&\int_0^T{\phi_{S,a}\mathcal{R}dt}+\phi^C_{Q,s}\\
    &=\int_0^T{\gamma(t)\phi_{\rm SF}^Ft\mathcal{R}(t)dt}+\phi^C_{Q,s}\\
    &\approx\gamma\phi_{\rm SF}^F\int_0^T{t\mathcal{R}(t)dt}+\phi^C_{Q,s}
\end{align}
Now we assume the fraction of self-quenched galaxies is the same in the cluster and field, so
\begin{equation}
    \frac{\phi^C_{\rm Q,s}}{\phi^C}=\frac{\phi^F_{\rm Q}}{\phi^F}
\end{equation}
and
\begin{equation}
    \phi_Q^C\approx\gamma\phi_{\rm SF}^F\int_0^T{t\mathcal{R}(t)dt}+\phi^F_{\rm Q}\frac{\phi^C}{\phi^F}.
\end{equation}
Divide both sides by $\phi^F$ to get
\begin{equation}
    \frac{\phi_Q^C}{\phi^F}\approx\gamma f_{\rm SF}^F\int_0^T{t\mathcal{R}(t)dt}+f_Q^F\frac{\phi^C}{\phi^F}.
\end{equation}
Using $\gamma$ from Equation~\ref{eq-gamma}, we obtain
\begin{equation}
    \frac{\phi_Q^C}{\phi^F}\approx\frac{1}{T}\frac{\phi^C}{\phi^F} f_{\rm SF}^F\int_0^T{t\mathcal{R}(t)dt}+f_Q^F\frac{\phi^C}{\phi^F} 
\end{equation}
or
\begin{equation}
    f_Q^C\approx\frac{1}{T} f_{\rm SF}^F\int_0^T{t\mathcal{R}(t)dt}+f_Q^F.
\end{equation}
This can be rearranged to give the QFE:
\begin{equation}\label{eq-QFEint}
    {\mbox QFE}=\frac{f_Q^C-f_Q^F}{f_{\rm SF}^F}\approx\frac{1}{T}\int_0^T{t\mathcal{R}(t)dt}.
\end{equation}
  In this case, a constant $QFE$ means $\mathcal{R}\propto t^{-1}$ and 
  \begin{equation}
      \mathcal{R}\approx\frac{QFE}{T}.
  \end{equation}
On the other hand, if the quenching rate $\mathcal{R}$ is constant, then 
\begin{equation}
    QFE\approx\frac{\mathcal{R}T}{2}
\end{equation}
should grow with time.  
This yields a simple, and intuitive relationship between $QFE$ and $\mathcal{R}$, but we end this section with a reminder that this depends on a number of strong assumptions that are likely to be incorrect in detail, including:
\begin{itemize}
    \item $\gamma$ and $\phi_{\rm SF}^F$ are both independent of time;
    \item $\mathcal{R}$ depends only on the {\it final} mass of a galaxy, and on time;
    \item Environmental quenching includes no change in stellar mass.  
\end{itemize}

\subsubsection{Constructing the Quiescent SMF from a Quenched Fraction Excess}\label{app-QFE_fM}
We note that when the cluster and field SMFs are not identical, the QFE cannot be simply interpreted as the fraction of field galaxies that have to be quenched and added to the field quiescent population in order to reproduce the cluster quiescent population.  That is,
\begin{equation}
    \phi_Q^C\neq A\left[\phi_Q^F+{(QFE)}\phi_S^F\right],
\end{equation}
where $A=\sum{\phi^C}/\sum{\phi^F}$ is the mass-independent renormalization constant.
Rather, from a simple rearrangement of terms in the definition of QFE,
\begin{equation}\label{eq-SMF_QFE}
    \phi_Q^C=\frac{\phi^C}{\phi_F}\left[\phi_Q^F+{(QFE)}\phi_S^F\right].
\end{equation}
Since the cluster and field total SMF do not have the same shape, the ratio outside the brackets is not independent of mass.
This is why, even though QFE is a weak function of stellar mass in GOGREEN clusters, the observed $\phi_Q^C$ cannot be reproduced by adding a fixed fraction of field SF galaxies to the field quiescent population, as demonstrated in \citet{van2020gogreen}.

It is useful to consider what function $f(M)$ {\it would} be required so that
\begin{equation}
    \phi_Q^C(M)=A\left[\phi_Q^F(M)+f(M)\phi_S^F(M)\right],
\end{equation}
given the constraints on QFE.  Here and for what follows for this subsection, we reintroduce the explicit $M$ dependence in the equations.  Combining Equations~\ref{eq-def_f} and \ref{eq-SMF_QFE} we solve for $f(M)$:
\begin{align}
    f(M)&=\frac{\phi_Q^C(M)/A-\phi_Q^F(M)}{\phi_S^F(M)}\\
     &=\frac{1}{A}\frac{\phi^C(M)}{\phi^F(M)}\left[\frac{\phi_Q^F(M)}{\phi_S^F(M)}+QFE(M)\right]-\frac{\phi_Q^F(M)}{\phi_S^F(M)}\\
     &=\xi(M)QFE(M)+\frac{\phi_Q^F(M)}{\phi_S^F(M)}\left[\xi(M)-1\right],\label{eq-f_QFE}
\end{align}
where we have defined
\begin{equation}
    \xi(M)=\frac{1}{A}\frac{\phi^C(M)}{\phi^F(M)}
\end{equation}
for convenience.  Note that, if the total SMF shape is independent of environment, then $\xi(M)=\frac{1}{A}\frac{\phi^C}{\phi^F}=1$ and Equation~\ref{eq-f_QFE} reduces to $f=QFE$.  Also interesting is that $f\neq 0$ when $QFE=0$.  That is, even if the quenched fraction is identical in the cluster and the field, it is still necessary to add quenched galaxies (and star forming galaxies) on top of what would be expected from the field in a mass dependent way, to account for the different total SMF shape.

\subsubsection{Transition galaxies}\label{app-trans}
\begin{figure}
    \centering
    \includegraphics[scale=0.55]{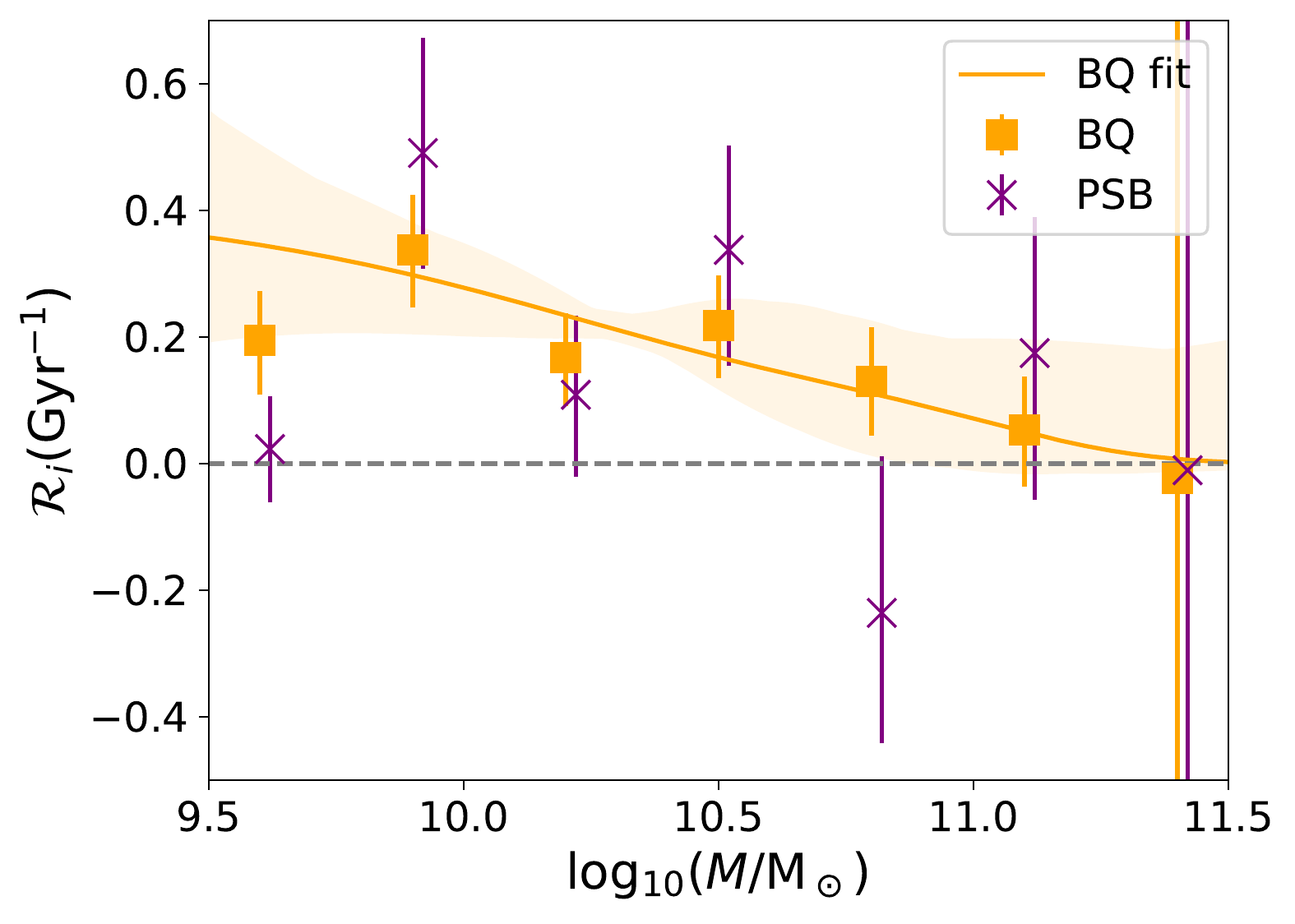}
    \caption{Similar to Figure~\ref{fig-Ratebqg}, we show the transition rate $\mathcal{R}$ for the BQ and PSB populations, but here assuming that the non-environmentally associated transition galaxy abundances are proportional to the SF population, rather than the total.  This increases the resulting rates, particularly at high stellar masses.  However, the change is modest and does not qualitatively affect our conclusions.}
    \label{fig-altRatebqg}
\end{figure}

Now consider what we learn from the excess abundance of transition galaxies, $f_x^C-f_x^F$.  Here $x$ can refer to PSB, BQ, GV or any other transition population.  
The procedure is similar to that in \S~\ref{app-QFE_R}.  We will assume that any SF galaxy accreted between time $T-\tau$ and $T$ will be observable as a transition galaxy at time $T$, given a lifetime of that phase, $\tau$.  
Therefore we have
\begin{align}
    \phi^C_x\approx&\int_{T-\tau}^T{\gamma\phi_S^F t\mathcal{R}(t)dt}+\phi_{x,s}^C\label{eq-trans_IE}\\
    &\approx\gamma\phi_S^F\int_{T-\tau}^T{t\mathcal{R}(t)dt}+\phi_{x,s}^C\\
    &\approx\gamma\phi_S^FT\tau\mathcal{R}(T)+\phi_{x,s}^C,
\end{align}
where in the last step we have assumed $\mathcal{R}$ is constant over the timescale $\tau\ll T$.
The additive term $\phi_{,s}^C$ is there to account for galaxies that meet the transition definition independently of environment, just like in Equation~\ref{eq-phiQC}. These include all galaxies classified as type $t$, whatever the (non-environmental) cause.  Recall that the term $\gamma\phi_S^F T$ in the integrand of Equation~\ref{eq-trans_IE} is just equal to $f_S^F\phi^C$; that is, it is the number of SF cluster galaxies that would be observed at time $T$ in the absence of environmental quenching.  
We can evaluate $\phi^C_{x,s}$ using the field,
\begin{equation}
    \frac{\phi^C_{\rm x,s}}{\phi^C_{\rm ref}}=\frac{\phi^F_x}{\phi^F_{\rm ref}},
\end{equation}
where we have written the abundances relative to some reference population $\phi_{\rm ref}$ in the field.  The choice of reference is not obvious here.  It may be appropriate to assume $\phi^j_{\rm ref}=\phi^j$ if the transition population in the field arises primarily from sources other than quenching of SF galaxies -- for example rejuvenation of Q galaxies, or scatter due to both physical causes (e.g. dust) or measurement uncertainties.  On the other hand, if the field population primarily reflects the quenching of SF galaxies it would be more correct to assume $\phi^j_{\rm ref}=\phi_{\rm SF}^j$.  Leaving this general for now, we obtain

\begin{equation}
    \phi^C_x\approx\gamma\phi_S^FT\tau\mathcal{R}(T)+\phi^F_x\frac{\phi^C_{\rm ref}}{\phi^F_{\rm ref}}.
\end{equation}
Dividing both sides by $\phi^F$ yields:

\begin{equation}
    \frac{\phi^C_x}{\phi^F}\approx\gamma f_S^FT\tau\mathcal{R}(T)+f^F_x\frac{\phi^C_{\rm ref}}{\phi^F_{\rm ref}}
\end{equation}
or 
\begin{equation}
    \frac{\phi^C}{\phi_F}f_x^C\approx\gamma f_S^FT\tau\mathcal{R}(T)+f^F_x\frac{\phi^C_{\rm ref}}{\phi^F_{\rm ref}}.
\end{equation}
Using Equation~\ref{eq-gamma} for $\gamma$,

\begin{equation}
    f_x^C\approx f_S^F\tau\mathcal{R}(T)+f^F_x\frac{f^C_{\rm ref}}{f^F_{\rm ref}}.
\end{equation}
Finally, we can write an expression for $\mathcal{R}$:
\begin{equation}
    \tau\mathcal{R}(T)=\frac{f_x^C-f^F_xf^C_{\rm ref}/f^F_{\rm ref}}{f_S^F}.
\end{equation}
This gives rise to Equation~\ref{eq-Rtau}, if the reference sample is the total sample so e.g. $f^C_{\rm ref}=\phi^C_{\rm ref}/\phi^C=1$.  We use this as our baseline, and the figures in the paper are generated under that assumption.  Alternatively, we might assume that the abundance of non-environmentally related transition galaxies are proportional to the fraction of star-forming galaxies in a population, so $f_{\rm ref}=f_{SF}$.  This has the effect of increasing the excess found in clusters, and hence the resulting $\mathcal{R}\tau$.  In Figure~\ref{fig-altRatebqg} we show the equivalent of Figure~\ref{fig-Ratebqg}, but under this assumption.  The difference is greatest at high stellar masses (where the ratio $f_S^C/f_S^F$ is smallest), increasing the inferred rate and decreasing the uncertainties.  The smaller uncertainties only strengthen our conclusion that the highest mass galaxies are consistent with an environmental quenching rate of $\mathcal{R}=0$.

\subsection{Quenching a fixed fraction of galaxies}
Alternatively, we can consider a model where a fixed fraction $\mathcal{Q}(M,t)$ of accreted galaxies is quenched.  This could correspond to a case where only a fraction of galaxies are on the right orbits to be quenched - and they either are or aren't, in a binary way.  Equation~\ref{eq-phi_Sa} becomes
\begin{align}\label{eq-Q}
    \phi^C_Q=&\int_0^T{\gamma\phi_S^F\mathcal{Q}dt}+\phi^C_{Q,s}\\
    &\approx\gamma\phi_S^F\int_0^T{\mathcal{Q}dt}+\phi^C_{Q,s}
\end{align}
Following through with the same algebra as in \S~\ref{app-QFE_R}, we end up with
\begin{equation}
    {\mbox QFE}=\frac{f_Q^C-f_Q^F}{f_S^F}\approx\frac{1}{T}\int_0^T{\mathcal{Q}(t)dt}.
\end{equation}
If $\mathcal{Q}$ is constant, then $QFE=\mathcal{Q}$.  This would be perhaps the simplest way to get a QFE that is independent of redshift and stellar mass.

\subsection{Primordial population}
Finally, let's consider the case where there is no accretion-based quenching at all, but the cluster simply starts with a population of "primordially quenched" galaxies that remains fixed in time.

\begin{align}
    \phi_Q^C=&\phi^C_{Q,p}+\phi^C_{Q,s}\\
     =&\phi^C_{Q,p}+\frac{\phi_Q^F}{\phi^F}\phi^C\\
     =&\phi^C_{Q,p}+f_Q^F\phi^C
\end{align}
Dividing both sides by $\phi^C$ then gives
\begin{equation}
    f_Q^C=\frac{\phi^C_{Q,p}}{\phi^C}+f_Q^F.
\end{equation}
So QFE is simply
\begin{equation}
    QFE=\frac{f_Q^C-f_Q^F}{f_S^F}=\frac{f^C_{Q,p}}{f_S^F}.
\end{equation}
Note that $f_{Q,p}^C$ will decrease with time, since (we assume) the primordial population $\phi_{Q,p}$ remains constant while the cluster $\phi^C$ grows around it.  
This can lead to a roughly constant QFE, since both $f^C_{Q,p}$ and $f_S^F$ decrease slowly with time.  However, it would not be natural to expect it to be constant with $M$, since $f_{Q,p}$ presumably increases with $M$, while $f_S^F$ decreases.

\section*{Affiliations}

$^{1}$Department of Physics and Astronomy, University of Waterloo, Waterloo, Ontario N2L 3G1, Canada \\
$^{2}$Waterloo Centre for Astrophysics, University of Waterloo, Waterloo, Ontario, N2L3G1, Canada \\
$^{3}$European Southern Observatory, Karl-Schwarzschild-Str. 2, 85748, Garching, Germany \\
$^{4}$INAF - Osservatorio astronomico di Padova, Vicolo Osservatorio 5, IT-35122 Padova, Italy \\
$^{5}$Department of Physics and Astronomy, The University of Kansas, 1251 Wescoe Hall Drive, Lawrence, KS 66045, USA \\
$^{6}$Department of Physics and Astronomy, York University, 4700 Keele Street, Toronto, Ontario, ON MJ3 1P3, Canada \\
$^{7}$Department of Physics and Astronomy, University of California, Irvine, 4129 Frederick Reines Hall, Irvine, CA 92697, USA \\
$^{8}$School of Physics and Astronomy, University of Birmingham, Edgbaston, Birmingham B15 2TT, England \\
$^{9}$INAF - Osservatorio Astronomico di Trieste, via G. B. Tiepolo 11, I-34143 Trieste, Italy \\
$^{10}$IFPU - Institute for Fundamental Physics of the Universe, via Beirut 2, 34014 Trieste, Italy \\
$^{11}$Departamento de Astronom\'ia, Facultad de Ciencias F\'isicas y Matem\'aticas, Universidad de Concepci\'on, Concepci\'on, Chile \\
$^{12}$Department of Physics and Astronomy, University of California, Riverside, 900 University Avenue, Riverside, CA 92521, USA \\
$^{13}$INAF-Osservatorio Astronomico di Trieste, via G. B. Tiepolo 11, 34143, Trieste, Italy\\
$^{14}$Departamento de Ingenier\'ia Inform\'atica y Ciencias de la Computaci\'on, Universidad de Concepci\'on, Concepci\'on, Chile \\
$^{15}$Department of Physics, University of Helsinki, Gustaf H\"allstr\"omin katu 2a, FI-00014 Helsinki, Finland \\
$^{16}$Laboratoire d'astrophysique, \'Ecole Polytechnique F\'ed\'erale de Lausanne (EPFL), 1290 Sauverny, Switzerland  \\
$^{17}$GEPI, Observatoire de Paris, Universit\'e PSL, CNRS, Place Jules Janssen, F-92190 Meudon, France \\
$^{18}$Research School of Astronomy and Astrophysics, The Australian National University, ACT 2601, Australia\\
$^{19}$ Centre for Gravitational Astrophysics, College of Science, The Australian National University, ACT 2601, Australia\\
$^{20}$Departamento de Ciencias F\'isicas, Universidad Andres Bello, Fernandez Concha 700, Las Condes 7591538, Santiago, Regi\'on Metropolitana, Chile\\
$^{21}$European Space Agency (ESA), European Space Astronomy Centre, Villanueva de la Ca\~{n}ada, E-28691 Madrid, Spain\\
$^{22}$Centro de Estudios de Física del Cosmos de Aragón, San Juan, 1 Planta-2, 44001 Teruel,  SPAIN\\
$^{23}$Department of Astronomy \& Astrophysics, University of Toronto, Toronto, Canada \\
$^{24}$Steward Observatory and Department of Astronomy, University of Arizona, Tucson, AZ 85721, USA \\

\bsp	
\label{lastpage}
\end{document}